\documentclass[fleqn,usenatbib]{mnras}

\usepackage{newtxtext,newtxmath}
\usepackage{ulem}

\usepackage[T1]{fontenc}
\usepackage{setspace}
\usepackage{enumitem}
\DeclareRobustCommand{\VAN}[3]{#2}
\let\VANthebibliography\thebibliography
\def\thebibliography{\DeclareRobustCommand{\VAN}[3]{##3}\VANthebibliography}

\usepackage{graphicx}	
\usepackage{amsmath}	

\usepackage{xcolor}

\definecolor{darkblue}{rgb}{0.0,0.0,0.3}
\definecolor{mediumgray}{gray}{0.5}
\definecolor{lightgray}{gray}{0.75}
\definecolor{verylightgray}{gray}{0.95}

\definecolor{blazeorange}{rgb}{1.0, 0.4, 0.0}
\definecolor{seagreen}{rgb}{0.18, 0.55, 0.34}
\definecolor{medgreen}{rgb}{0,0.8,0}
\definecolor{rufous}{rgb}{0.66, 0.11, 0.03}
\definecolor{royalfuchsia}{rgb}{0.79, 0.17, 0.57}
\definecolor{scarlet}{rgb}{1.0, 0.13, 0.0}
\definecolor{royalpurple}{rgb}{0.47, 0.32, 0.66}
\definecolor{orange}{rgb}{1.0, 0.5, 0.5}

\newcommand\MJs[0]{MJs/patches}

\newcommand{\fracb}[2]{\left(\frac{#1}{#2}\right)}

\title[Polarization from Non-Axisymmetric Jets]{Prompt GRB Polarization from Non-Axisymmetric Jets}

\author[Gill \& Granot]{
Ramandeep Gill$^{1,2}$\thanks{E-mail: r.gill@irya.unam.mx}
and Jonathan Granot$^{2,3,4}$\thanks{E-mail: granot@openu.ac.il}
\\
$^{1}$Instituto de Radioastronomía y Astrofísica, Universidad Nacional Autónoma de México, Antigua Carretera a Pátzcuaro \# 8701, \\
Ex-Hda. San José de la Huerta, Morelia, Michoacán, México C.P. 58089 \\
$^{2}$Astrophysics Research Center of the Open university (ARCO), The Open University of Israel, P.O Box 808, Ra'anana 4353701, Israel\\
$^{3}$Department of Natural Sciences, The Open University of Israel, P.O Box 808, Ra'anana 4353701, Israel\\
$^{4}$Department of Physics, The George Washington University, Washington, DC 20052, USA
}

\date{Accepted XXX. Received YYY; in original form ZZZ}

\pubyear{2023}

\begin{document}
\label{firstpage}
\pagerange{\pageref{firstpage}--\pageref{lastpage}}
\maketitle

\begin{abstract}
Time-resolved linear polarization ($\Pi$) measurements of the prompt gamma-ray burst 
emission can reveal its dominant radiation mechanism. A widely considered mechanism is 
synchrotron radiation, for which linear polarization can be used to probe the 
jet's magnetic-field structure, and in turn its 
composition. In axisymmetric jet models the polarization angle (PA) can only 
change by $90^\circ$, as $\Pi$ temporarily vanishes. However, some time-resolved 
measurements find a continuously changing PA, which requires the flow to be non-axisymmetric 
in at least one out of its emissivity, bulk Lorentz factor or magnetic field.
Here we consider synchrotron emission in non-axisymmetric jets, from an ultrarelativistic 
thin shell, comprising multiple radially-expanding mini-jets (MJs) or emissivity 
patches within the global jet, that yield a continuously changing PA. We explore a wide 
variety of possibilities with emission consisting of a single pulse or multiple overlapping 
pulses, presenting time- resolved and integrated polarization from  different magnetic 
field configurations and jet angular structures. We find that emission from multiple incoherent MJs/patches 
reduces the net polarization due to partial cancellation in the Stokes plane. When these contain 
a large-scale ordered field in the plane transverse to the radial direction, $\Pi$ always starts 
near maximal and then declines over the single pulse or shows multiple highly polarized peaks 
due to multiple pulses. Observing $\Pi\lesssim40\%$ (15\%) integrated over one (several) pulse(s) 
will instead favor a shock-produced small-scale field either ordered in the radial direction or tangled in the plane transverse to it.
\end{abstract}




\section{Introduction}

Despite many efforts over the last several decades, the exact radiation mechanism that produces the Band-like \citep{Band+93} 
non-thermal spectrum of prompt emission in gamma-ray bursts (GRBs) remains poorly understood 
\citep[see, e.g.,][for a review]{Kumar-Zhang-15}. The two most favored mechanisms are 
optically-thin synchrotron emission from relativistic electrons with power-law energy distribution 
\citep[e.g.][]{Sari-Piran-97,Daigne-Mochkovitch-98,Genet-Granot-09,RGB23} 
and inverse-Compton scattering of softer seed quasi-thermal photons 
\citep[e.g.][]{Thompson-94,Ghisellini-Celotti-99,Giannios-06,Thompson-Gill-14,Gill-Thompson-14}. 
Spectral modeling of prompt GRB emission alone has proven insufficient in discriminating between these two radiation mechanisms 
\citep[e.g.][]{Gill+20b}. One promising tool that could break this degeneracy is linear polarization \citep[see, e.g.,][for a review]{Gill+21}. 

Linear polarization can also be used to better understand our viewing geometry, the jet's angular structure, and for 
synchrotron emission also the jet's B-field configuration. Time-integrated \citep{Granot-03,Lyutikov-Blandford-03,Toma+09,Gill+20} and 
time-resolved \citep{Gill-Granot-21} polarization models for synchrotron emission from \textit{axisymmetric} outflows have been presented in many works. 
In this case, due to the azimuthal symmetry of the emissivity, bulk Lorentz factor (LF) $\Gamma$ and the B-field in the emitting region, 
the net polarization angle (PA) $\theta_\Pi$ can only align with two directions, either along the 
line connecting the observer's line-of-sight (LOS) to the jet symmetry axis or transverse to it, regardless of the emission mechanism. 
Consequently, the change in PA can only be $\Delta\theta_\Pi=90^\circ$. One way this symmetry can be broken is when the observed 
region of angular size $1/\Gamma$ around our LOS contains an ordered B-field. That can 
lead to a gradually and continuously changing PA \citep{Granot-Konigl-03,Wang-Lan-2023a,Wang-Lan-2023b}.

Time-resolved polarization measurements in GRB\,170114A by POLAR did hint at a continuously evolving PA \citep{Zhang+19,Burgess+19}. 
Similar changes, albeit integrated over larger time-bins, have also been seen (with modest significance) in some other GRBs \citep{Yonetoku+11,Chand+19,Sharma+19}. 
Axisymmetric jet models cannot explain a continuously changing PA, and hence the need to develop \textit{non-axisymmetric} models 
of prompt GRB polarization. Non-axisymmetric jet models were discussed in many earlier works 
\citep[e.g.,][]{Shaviv-Dar-95,Gruzinov-Waxman-99,Granot-Konigl-03,Lyutikov-Blandford-03,Nakar-Oren-04,Lyutikov-06,Lazar+09,Narayan-Kumar-09,Zhang-Yan-11,Huang-Liu-22}.
In many cases this was in the context of GRB afterglows, where the emission at any given observed time originates 
from several \textit{mini-jets} (MJs) or \textit{patches} within a larger collimated outflow or global jet. 
Such patches or MJs can vary in emissivity, bulk-$\Gamma$, and/or B-field structure according to some distribution as a function 
of polar angle $\theta$ and azimuthal angle $\varphi$ measured, respectively, from and around the jet symmetry axis.

In this work we generalize a model of time-resolved prompt GRB polarization introduced in \citet{Gill-Granot-21} to 
include several MJs\,/\,patches, and demonstrate how a continuously varying PA can be obtained for different jet 
structures and  B-field configurations when the emission mechanism 
is synchrotron radiation. The MJs and patchy shell models are discussed in \S\ref{sec:MJ-PS}. The formalism for calculating 
time- and energy-dependent pulse profiles and linear polarization from an axisymmetric ultrarelativistic outflow is presented in \S\ref{sec:theory-model}. 
In \S\ref{sec:non-axisymmetric-jets}, we present our model of non-axisymmetric jets including MJs or patches and calculate the polarization evolution 
over a single pulse. In \S\ref{sec:multi-pulse} the single pulse formalism is used to obtain the polarization properties of multiple overlapping pulses. Our conclusions are given in \S\ref{sec:conclusion-discussion} along with a discussion of existing time-resolved polarimetric observations and their 
interpretation within our model. In Appendix\;\ref{sec:appendix} we present results for several additional cases to capture 
the variety in polarization evolution expected in the scenarios explored here.

 \section{Mini-Jets and Patchy Shells}\label{sec:MJ-PS}
Multiple mutually-incoherent emitting regions can arise within the aperture of the global jet because of
some kind of hydrodynamic or hydromagnetic disturbance. The most likely nature of this disturbance depends 
on the outflow composition (see \S\ref{sec:composition-dynamics} below), i.e. whether it is kinetic-energy-dominated 
(i.e. weakly magnetized) in which case it is ascribed to internal shocks between different shells launched by 
the central engine, or Poynting-flux-dominated (i.e. strongly magnetized) where a magnetohydrodynamic (MHD) instability can produce it. In both 
cases, the distinct emission regions can be envisaged either as individual blobs or MJs with mutually distinct 
properties or patches that are assumed (for simplicity) to vary only in emissivity. 
The MJs can vary in their angular structure, e.g. uniform top-hat MJs or MJs with a core and power-law wings. 
The distinction between MJs and patches disappears when the global jet is uniform in bulk-$\Gamma$ and the MJs have a top-hat 
angular profile. 

In general, the emitting blobs can be moving into random directions with a distribution of LFs $\gamma_b'$ in the 
mean outflow (local center of momentum) rest frame, which itself is moving with bulk LF $\Gamma\gg\gamma_b'$ 
\citep{Lyutikov-Blandford-03,Lyutikov-06,Narayan-Kumar-09,Lazar+09,Beniamini-Granot-16,Granot-16}. 
As a result, a given observer will only receive emission from a fraction of the total number of such blobs whose beaming cones 
(of angular size $\sim\!1/\Gamma\gamma_b'$ in the lab-frame) point towards the observer. This also implies that emission from multiple blobs 
can be observed within the $1/\Gamma$ beaming cone of the outflow around the LOS.

Here we adopt a simpler scenario \citep{Kumar-Piran-00}, where we ignore the random motions of the emitting regions away from the radial 
direction in the fluid frame, and instead assume that the MJs and patches propagate radially. 
The angular size of the emitting regions, $\bar\theta\sim1/\Gamma$, is dictated by causality, 
where $\Gamma$ is the local bulk Lorentz factor of the flow. Regions separated by angular scales much larger than $1/\Gamma$ must 
remain causally disconnected. Therefore, the total number of MJs or patches is limited to, e.g., $N\sim(\Gamma\theta_j)^2$ for a top-hat 
global jet with half-opening angle $\theta_j$. Since the observed angular size of the flow, set by the beaming cone, is also $\sim1/\Gamma$, 
no more than a single MJ or patch can be observed fully from within its beaming cone. However, multiple such regions can be partially 
observed, from outside of their beaming cones, depending on their 
distribution and covering factor. Contribution to the observed emission from neighbouring patches that lie outside of the beaming 
cone, as shown in later sections, must arrive later due to the angular time delay and must also be suppressed due to Doppler de-beaming.

Prompt GRB emission shows strong temporal variability on a timescale $t_v\ll t_{\rm GRB}$ where $t_{\rm GRB}$ is the 
total duration of the prompt emission. This can be readily attributed to multiple shell 
collisions in the internal shock scenario, but in general may require more elaborate models, e.g. relativistic turbulence \citep{Lazar+09}
or magnetic reconnection \citep{Lyutikov-06}.

\subsection{Magnetic Field}\label{sec:B-field}

To calculate the linear polarization, we need to know the structure of the magnetic field in the emission region. Here 
we consider four physically motivated structures \citep[e.g.,][]{Gill+20}: (i) $B_{\rm ord}$: an ordered B-field in the 
plane transverse to the radial direction and having a coherence length angular scale larger than that of the beaming cone, such that 
$\theta_B\gtrsim\Gamma^{-1}$ \citep{Granot-03}; (ii) $B_\perp$: a shock-generated tangled (randomly oriented) B-field with 
$\theta_B\ll\Gamma^{-1}$ also constrained to be in the plane transverse to the radial direction \citep{Granot-03}; 
(iii) $B_\parallel$: an alternative to the previous case and a generalization of the shock-generated field, where the field 
is now ordered in the radial direction \citep{Granot-03}; (iv) $B_{\rm tor}$: an axisymmetric globally ordered toroidal field that 
is expected to arise in PFD outflows \citep{Lyutikov+03,Granot-Taylor-05}. Please note that apart from $B_{\rm ord}$, the 
other field structures possess global axisymmetry w.r.t the jet axis.

Large-scale ordered B-field configurations, such as $B_{\rm ord}$ and $B_{\rm tor}$, persist in Poynting-flux-dominated outflows. 
Near the jet launching radius the B-field is anchored either in the accretion disk and/or a rapidly rotating magnetar \citep[e.g.][]{Spruit+01}. 
Since the poloidal component of an axisymmetric field declines much more rapidly with radius ($B_p\propto R^{-2}$) in comparison to 
the toroidal component ($B_\varphi\propto R^{-1}$), the field at large distances is predominantly transverse to the radial direction. 
It is also susceptible to current-driven magnetic kink instabilities and/or turbulence at the interface between the jet and confining 
medium \citep[e.g.][]{Bromberg-Tchekhovskoy-16,Lazarian+19}, that distort its shape and introduce random polarity reversals in different directions across the jet surface (see, e.g., Fig.\,4 of \citealt{Mckinney-Uzdensky-12} and Fig.\,2 of \citealt{Kadowaki+21}). 
As a result, the field reconnects at different locations within the bulk flow and produces jets of relativistically moving 
electrons into random directions. These electrons then radiate synchrotron photons into a cone of angular size $1/\gamma_b'$ around 
their direction of motion in the bulk flow. The complexity and appropriate physics of such a scenario is not captured by the simpler model of radially propagating 
blobs/MJs explored in this work. The coherence of the non-reconnecting magnetic field, along which the electrons propagate and 
cool, is maintained on angular scales $\theta_B\gtrsim1/\Gamma$, and these large scale fields occupy a large fraction of the jet aperture.  
In fact, the presence of dynamically-dominant large-scale fields would actually suppress the formation of blob-like 
structures in the flow as that would require  
severely bending field lines against magnetic tension, which is challenging to achieve in a strongly magnetized outflow.  
An additional source of free energy, such as magnetic reconnection or proper velocity variations resulting 
in internal collisions and in turn some magnetic reconnection, may lead to the formation of local small blobs, but 
these are typically expected on angular scales $\ll1/\Gamma$.
Therefore, 
the simpler MJ scenario assumed here is not so well physically motivated in this case. Instead, we describe the non-axisymmetric emission 
in this case using a patchy shell with distinct emissivity patches. These represent the locations where the accelerated electrons lose most of their energy to synchrotron radiation while propagating along the non-reconnecting large-scale fields. In the $B_{\rm tor}$ scenario, it is further assumed 
that the non-reconnecting field retains its initial toroidal structure even after some of the field reconnects locally due to instabilities in 
the flow.

On the other hand, the MJ scenario is well suited to describe a kinetic-energy-dominated outflow, in which collisions of inhomogeneous and 
radially propagating shells may in turn produce several radially propagating blobs. These collisions would also lead to internal shocks 
and therefore shock-produced fields, such as $B_\perp$ or  
$B_\parallel$, that represent two extremes of the field anisotropy with respect to the direction of the local shock normal, 
which aligns with the lab frame fluid velocity (assumed radial in this work). The true anisotropy of the field at such collisionless 
shocks is not entirely clear yet and it may be intermediate between these two extremes, and evolve with the distance behind the shock 
\citep{Granot-Konigl-03,Gill-Granot-20}.

\section{Axisymmetric Jet Polarization Model}\label{sec:theory-model}

We consider the dynamics and emission of an ultrarelativistic thin-shell, with LF $\Gamma\gg1$ and lab-frame width 
$\Delta\ll R/\Gamma^2$.  
The shell radiates between radii $R_0$ and $R_f=R_0+\Delta R$. During this time, its dynamical evolution is governed by 
$\Gamma(R)=\Gamma_0(R/R_0)^{-m/2}$, where $\Gamma_0=\Gamma(R_0)$. The index $m$ is used to study cases in which the shell is coasting ($m=0$), 
accelerating ($m<0$), or decelerating ($m>0$). All comoving quantities henceforth appear with a prime. 
The shell's comoving anisotropic (w.r.t. the local B-field direction) spectral luminosity evolves with radius as the 
peak luminosity and spectral peak energy change as a power law with radius,
\begin{equation}\label{eq:L-and-nupk-scaling}
    L'_{\nu'}(R,\theta) = L_0'\fracb{R}{R_0}^a  S\fracb{\nu'}{\nu'_{\rm pk}}\,f(\theta)\quad{\rm with}\quad\nu'_{\rm pk} = \nu_0'\fracb{R}{R_0}^d\,,
\end{equation}
where $L_0' = L'_{\nu_{\rm pk}'}(R_0)$ and $\nu_0' = \nu_{\rm pk}'(R_0)$ are normalizations of the spectral luminosity 
and peak frequency at $R=R_0$, and $f(\theta)$ encodes the jet (axisymmetric) angular structure. 
The comoving spectrum is assumed to be the phenomological Band-function \citep{Band+93},
\begin{equation}
    S(x) = e^{1+b_1} \left\{
    \begin{tabular}{c|c}
        $x^{b_1} e^{-(1+b_1)x}$ & $x\leq x_b$\;, \\
        $x^{b_2}x_b^{b_1-b_2}e^{-(b_1-b_2)}$ & $x\geq x_b$\;,
    \end{tabular}\right.
\end{equation}
where $x \equiv \nu'/\nu_{\rm pk}' = (2\Gamma_0/\delta_D)x_0(R/R_0)^{-d}$ with $x_0\equiv(\nu/\nu_0)$.  
Here $\nu_0=2\Gamma_0\nu'_0/(1+z)$ ($z$ being the source redshift) is the peak frequency of the first photons emitted along the observer's LOS 
from radius $R_0$ and received at time $t=t_0$, and $\delta_D$ is the Doppler factor defined below. 
The break energy $x_b = (b_1-b_2)/(1+b_1)>1$ when $b_2<-1$. 
The power-law indices $a$ and $d$ in Eq.\,(\ref{eq:L-and-nupk-scaling}) depend on the outflow composition 
and dynamics, which we briefly discuss next.

\begin{table}
    \centering
    \begin{tabular}{|c|l|}
        Symbol & Definition \\
        \hline
        $a$ & Comoving spectral luminosity \\
        & radial power-law (PL) index: $L_{\nu'}'\propto R^a$ \\
        $a_j$ & Global jet comoving peak luminosity \\
        & angular profile PL index \\
        $a_{\rm mj}$ & MJ/patch comoving peak luminosity \\
        & angular profile PL index \\
        $b_j$ & Global jet bulk-$\Gamma$ angular profile PL index \\
        $b_1$, $b_2$ & Asymptotic Band-function spectral indices: \\
                & $d\ln F_\nu/d\ln\nu$ \\
        $d$ & Comoving peak frequency PL index: $\nu_{\rm pk}'\propto R^d$ \\
        $\mathcal{F}$ & MJ/patch covering factor \\
        $\Gamma_c$ & Core bulk-$\Gamma$ in a structured jet \\
        $m$ & Bulk-$\Gamma$ radial PL index: $\Gamma^2\propto R^{-m}$ \\
        $N_{\rm pulse}$ & Number of overlapping pulses \\
        & in an emission episode \\
        $\nu_0$ & $\nu F_\nu$ peak frequency of first photons emitted \\
        & from $R_0$ along the LOS that arrived at time $t_0$\\
        $\Pi$ & Polarization degree \\
        $q$ & $\theta_{\rm obs}/\theta_j$ (Uniform global jet) \\
        & $\theta_{\rm obs}/\theta_c$ (Structured global jet) \\
        $R_0$ & Radius at which emission turns on \\
        $\Delta R$ & Radii over which shell emits continuously \\
        $t_0$ & Arrival time of first photons emitted \\
        & along the LOS at $R_0$ \\
        $\tilde t$ & Normalized observer time: $\tilde t = t/t_0$ \\
        $\theta_\Pi$ & Polarization angle \\
        $\theta_j$ & Half-opening angle of the global jet \\
        $\theta_c$ & Core angular size in a structured global jet \\ 
        $\theta_{\rm obs}$ & Observer's viewing angle \\
        $\bar\theta_{\rm patch/mj}$ & Half-opening angle of the patch/MJ \\
        $\bar\theta_{{\rm patch/mj},c}$ & $\bar\theta_{\rm patch/mj}$ at $\theta_{\rm patch/mj}=0$ in a \\
        & structured global jet \\
        $\{\theta,\varphi\}_{\rm patch/mj}$ & Polar and azimuthal coordinate \\
         & of patch/MJ symmetry axis \\
         $x_0$ & Normalized observation frequency ($\nu/\nu_0$) \\
        & for a uniform jet \\
        $x_{0,c}$ & $x_0$ for $\theta_{\rm obs}=0$ in a structured global jet \\
        $\xi_j$, $\xi_{\rm patch/mj}$ & Uniform global jet: $(\Gamma\theta_j)^2$, $(\Gamma\bar\theta_{\rm patch/mj})^2$ \\
        $\xi_c$, $\xi_{{\rm patch/mj},c}$ & Structured global jet: $(\Gamma_c\theta_c)^2$, $(\Gamma_c\bar\theta_{{\rm patch/mj},c})^2$ \\
        $\xi_{\rm mj,\min}$ & Minimum allowed $\xi_{\rm patch/mj}$ in a \\
        & structured global jet \\
        \hline
    \end{tabular}
    \caption{Most relevant symbols and their definitions; See Table\,1 in \citet{Gill-Granot-21} for a complete list.}
    \label{tab:symbols}
\end{table}

\subsection{Outflow Composition and Dynamics}\label{sec:composition-dynamics}
The outflow's dynamical evolution and how energy is dissipated and then ultimately radiated, 
are sensitive to its composition, typically expressed in terms of its magnetization 
$\sigma$, given by the ratio of the magnetic to matter enthalpy densities. When $\sigma\ll1$, the flow dynamics 
are described by the fireball model \citep{Rees-Meszaros-94,Paczynski-Xu-94,Sari-Piran-97,Daigne-Mochkovitch-98}, 
in which after an initial rapid acceleration phase with $\Gamma(R)\propto R$, the fireball LF saturates to 
$\Gamma_\infty\gtrsim100$ and the outflow becomes kinetic-energy dominated.
During this coasting phase ($m=0$) part of the kinetic energy is dissipated and radiated away in internal shocks. 

Alternatively, when $\sigma\gtrsim1$, the outflow is Poynting flux dominated \citep[e.g.,][]{Thompson-94,Lyutikov-Blandford-03},  
and the main energy reservoir is magnetic. This energy may be dissipated via magnetic reconnection and/or 
MHD instabilities, e.g. the Kruskal-Schwarzchild instability \citep{Lyubarsky-10,Gill+18} 
which is the magnetic analog of the Rayleigh-Taylor instability. These can develop in models of high-$\sigma$ outflows featuring a striped wind 
\citep{Lyubarsky-Kirk-01,Spruit+01,Drenkhahn-02,Drenkhahn-Spruit-02,Begue+17} in which the magnetic field lines 
reverse polarity and magnetic energy is dissipated when opposite polarity field lines undergo reconnection. 
A significant fraction of the dissipated energy goes towards accelerating the flow 
as $\Gamma\propto R^{1/3}$, i.e. with $m=-2/3$. A similar acceleration profile is also obtained in a highly time-variable 
high-$\sigma$ outflow even without field polarity reversals \citep{Granot+11,Granot-12,Komissarov12}, which may still dissipate 
energy via internal shocks.

\citet{Gill-Granot-21} showed that for a kinetic-energy-dominated (KED) flow the power-law indices in 
Eq.\,(\ref{eq:L-and-nupk-scaling}) are $a=1$ and $d=-1$ \citep[see also][]{Genet-Granot-09}, while these for the Poynting-flux-dominated model 
are $a=4/3$ and $d=-2$. For brevity, we only present results for the KED flow here and note that these can be 
easily generalized to different $m$ values.

\subsection{Lightcurve and Polarization}
The energy-dependent linear polarization, $\Pi_\nu=\sqrt{Q_\nu^2+U_\nu^2}\big/I_\nu$, can be expressed 
using the Stokes parameters $\{I_\nu,Q_\nu,U_\nu,V_\nu\}$ where $V_\nu=0$ (negligible circular polarization) is expected,
$I_\nu\propto F_\nu$ is the total specific intensity at frequency $\nu$, and $F_\nu$ is the flux density. 
The ratio of the Stokes parameters is given by \citep{Gill+20}, 
\begin{eqnarray}\label{eq:general-pol}
    \left\{\begin{aligned}
        \frac{Q_\nu(t_z)}{I_\nu(t_z)} \\
        \frac{U_\nu(t_z)}{I_\nu(t_z)}
    \end{aligned} \right\} 
    =& 
    \\ \nonumber
    &\hspace{-0.38cm}\frac{\int_{\hat R_{\min}}^{\hat R_{\max}} d\hat R \left\vert \frac{d\tilde\mu}{d\hat R}\right\vert \delta_D^3 L'_{\nu'}(\hat R) 
    \int_0^{2\pi}d\tilde\varphi \Lambda(\tilde\xi,\tilde\varphi)\Pi'
    \left\{\begin{aligned}
        \cos(2\theta_p) \\
        \sin(2\theta_p)
    \end{aligned}\right\}
    }
    {\int_{\hat R_{\min}}^{\hat R_{\max}} d\hat R \left\vert \frac{d\tilde\mu}{d\hat R}\right\vert \delta_D^3 L'_{\nu'}(\hat R) 
    \int_0^{2\pi}d\tilde\varphi \Lambda(\tilde\xi,\tilde\varphi)}\,,
\end{eqnarray}
for radiation received by a distant observer in the direction of the unit vector $\hat n$ from a thin-shell located at a redshift $z$ 
and expanding with LF $\Gamma=(1-\beta^2)^{-1/2}$ and radial velocity $\vec\beta c$, with $c$ being the speed of light. 
The Doppler factor relates the observed and comoving frequency of radiation emitted by material moving in the 
direction of the unit vector $\hat\beta=\vec\beta/\beta$, and it is given by 
$\delta_D=(1+z)\nu/\nu'=[\Gamma(1-\vec\beta\cdot\hat n)]^{-1}=[\Gamma(1-\beta\tilde\mu)]^{-1}$, 
where $\hat n\cdot\hat\beta\equiv\tilde\mu=\cos\tilde\theta$ with $\tilde{\theta}=\tilde{\theta}(\theta,\varphi)$ and 
$\tilde\varphi$ being the polar and azimuthal angles measured, respectively, from and around the LOS. 
For an ultrarelativistic flow $\Gamma\gg1$ and for which $\delta_D\approx2\Gamma/(1+\tilde\xi)$ with $\tilde\xi\equiv(\Gamma\tilde\theta)^2$. 
The arrival time $t$ of a photon originating at an angle $\tilde\theta$ and from a radius $R$ is given by \citep[e.g.][]{Granot+08}
\begin{equation}\label{eq:EATS}
    t_z(R,\tilde\mu)\equiv\frac{t}{(1+z)}=t_{\rm lab}-\frac{R}{c}\tilde\mu \approx \frac{R}{2\Gamma^2c}\left(\tilde\xi + \frac{1}{1+m}\right)\,,
\end{equation}
for an ultrarelativistic thin-shell expanding with bulk LF $\Gamma(R)\propto R^{-m/2}$ with $m>-1$, and where 
$t_{\rm lab} = \int_0^R dR'/c\beta(R')$ is the lab-frame time. Here we made the approximation $\tilde\mu\approx1-\tilde\theta^2/2$ 
for $\tilde\theta\ll1$ when $\Gamma\gg1$ and where $\tilde\xi = (\Gamma\tilde\theta)^2 = (\Gamma_0\tilde\theta)^2\hat R^{-m} = \tilde\xi_0\hat R^{-m}$ 
with $\hat R\equiv R/R_0$. The arrival time of the first photons originating at radius $R_0$ and along the LOS with 
$\tilde\mu=1~(\tilde\xi=0)$ is $t_{0,z}=R_0/2(1+m)\Gamma_0^2c$.

The comoving synchrotron spectral luminosity $L'_{\nu'}$ is anisotropic and $\Lambda(\tilde\xi,\tilde\varphi)=\langle[1-(\hat n'\cdot\hat B')^2]^{\epsilon/2}\rangle$ 
represents the factor relating to the pitch angle of radiating electrons averaged over the local probability distribution of the comoving magnetic field 
$\hat B'$ in Eq.\,(\ref{eq:general-pol}), and where $\Pi'$ and $\theta_p$ are the local (and not averaged over the whole observed region of the emitting shell) degree 
of polarization and position angle, respectively. When the power-law electrons' energy distribution is independent of their pitch angles, $\epsilon=1+\alpha$ where 
$\alpha=-d\log F_\nu/d\log\nu$ is the spectral index \citep{Laing-80,Granot-03}. Expressions for $\Lambda(\tilde\xi,\tilde\varphi)$, 
$\theta_p$, and $\Pi'$ for different magnetic field configurations and assuming an ultrarelativistic uniform flow were first derived in \citet{Granot-03,Granot-Konigl-03,Lyutikov+03,Granot-Taylor-05} and are summarized in \citet{Toma+09,Gill+20}.

\begin{figure}
    \centering
    \includegraphics[width=0.4\textwidth]{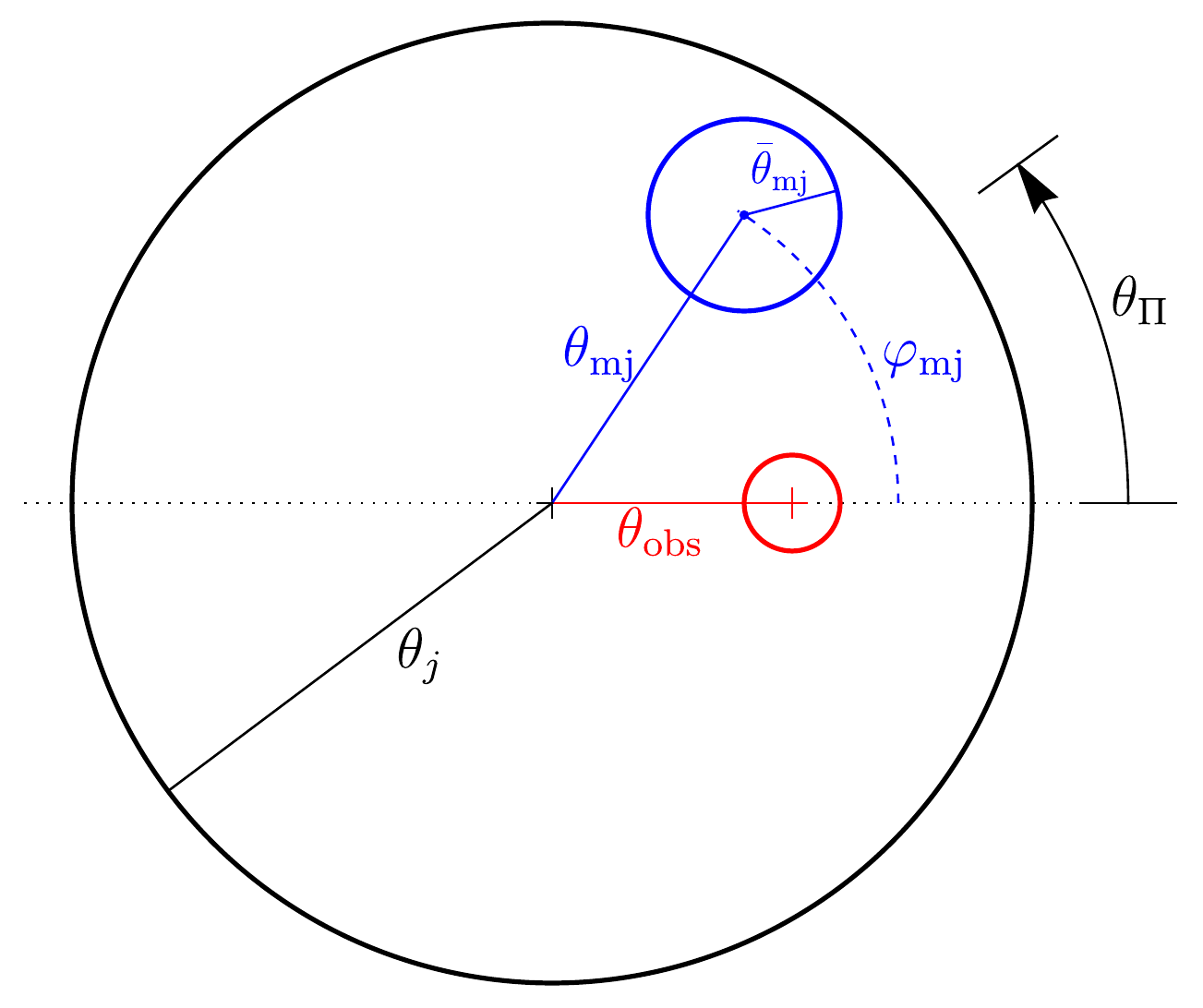}
    \caption{The physical setup showing different angular scales. A single top-hat MJ (blue) or a uniform patch, 
    with half-opening angle $\bar\theta_{\rm mj}$, is shown inside a top-hat global jet with angular size $\theta_j$. 
    The symmetry axis of the MJ has coordinates ($\theta_{\rm mj},\varphi_{\rm mj}$) w.r.t the global jet axis, 
    where the azimuthal angle (and the polarization angle $\theta_\Pi$) is measured counter-clockwise from the line 
    connecting the global jet symmetry axis and observer's line-of-sight (red plus sign). The red circle shows the 
    beaming cone of angular size $1/\Gamma$. The same setup is also generalized to global jets with angular structure 
    with $\theta_j\rightarrow\theta_c$, where $\theta_c$ is the angular size of the core.
    }
    \label{fig:geometry}
\end{figure}

\begin{figure*}
    \centering
    \includegraphics[width=0.23\textwidth]{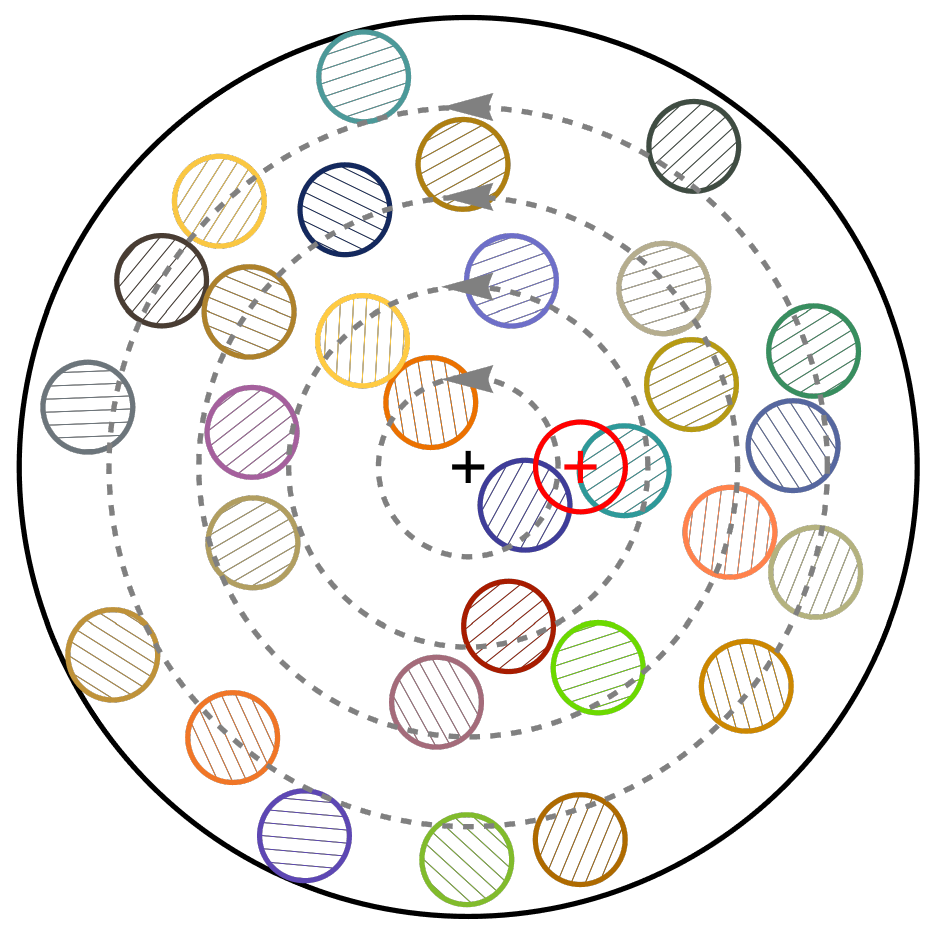} \\
    \centering
    \includegraphics[width=\textwidth]{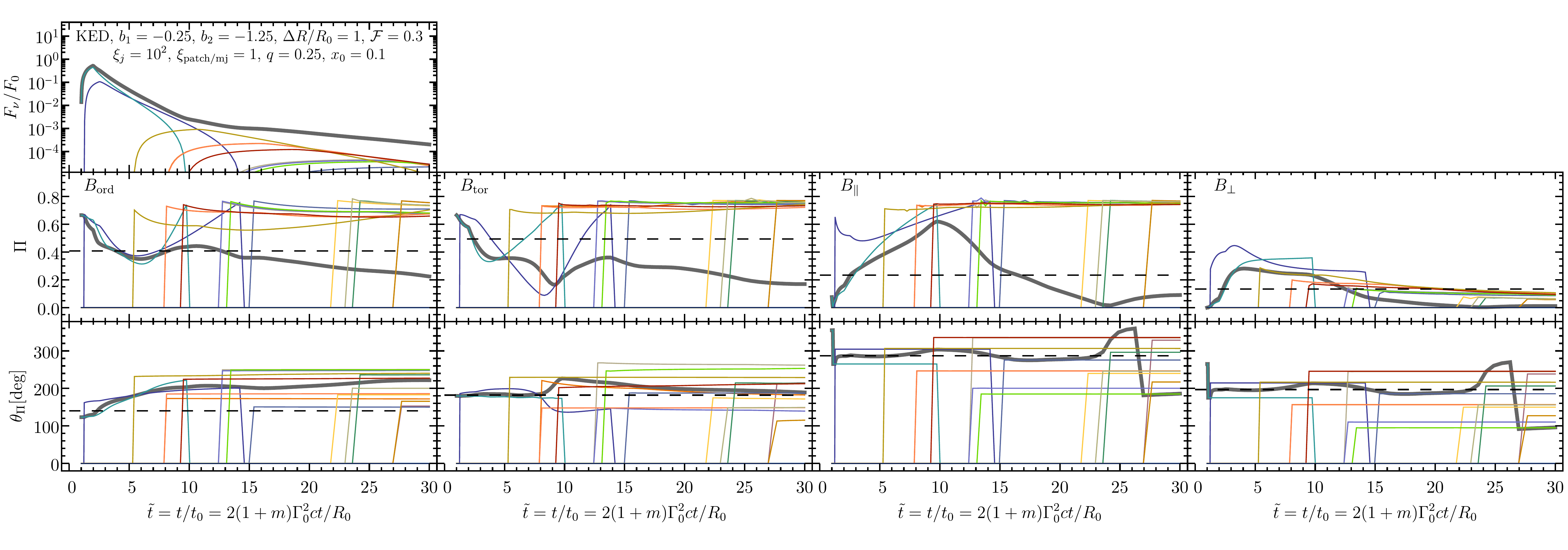}
    \caption{{\bf Top:} A global top-hat jet (solid black circle) comprising several top-hat MJs (used for $B_\parallel$ and $B_\perp$) 
    or uniform emissivity patches on the global jet's surface (used for $B_{\rm ord}$ and $B_{\rm tor}$). The global jet and MJs
    /patches have normalized (by the causal size) solid angles $\xi_j=(\Gamma\theta_j)^2=10^2$ and $\xi_{\rm patch/mj}=(\Gamma\bar\theta_{\rm patch/mj})^2=1$, 
    respectively, with a covering factor $\mathcal{F}=0.3$ implying $N_{\rm mj/patch}=30$ MJs/patches. 
    The $B_{\rm ord}$ field is illustrated here with the 
    patches having an ordered B-field with coherence angular scale of $\theta_B\sim\bar\theta_{\rm patch}$ in the plane 
    transverse to the radial direction. The globally ordered toroidal field $B_{\rm tor}$ is shown with gray dashed concentric circles around the 
    global jet's symmetry axis marked by a `+' symbol. The size of the beaming cone is shown with a red circle, having $\xi=(\Gamma\theta)^2=1$, around the observer's line-of-sight (LOS) marked by a red `+' symbol.  
    {\bf Bottom:} Pulse profile and polarization evolution over a single pulse for a kinetic-energy-dominated (KED)  
    flow with $\Gamma(R)=\Gamma(R_0)=\Gamma_0\gg1$. The flux density is normalized by the peak flux density ($F_0$) from a 
    MJ/patch centered at the LOS, and the time is normalized by the arrival time ($t_0$) of the first photons emitted along the LOS 
    by material at radius $R_0$. The contribution of the $N_{\rm mj/patch}$ distinct MJs/patches is shown with different 
    colors, while the global quantities are shown with a black curve. These are the pulse profile (top), linear 
    polarization (middle), and polarization angle (bottom) measured counter-clockwise from the horizontal axis. The black dashed line in the middle 
    panel shows the time-integrated polarization. Only a single lightcurve panel is shown as the differences in lightcurves for the different B-fields are negligibly small. }
    \label{fig:THmj-in-THJ-diff-Bfield}
\end{figure*}

\begin{figure*}
    \centering
    \includegraphics[width=\textwidth]{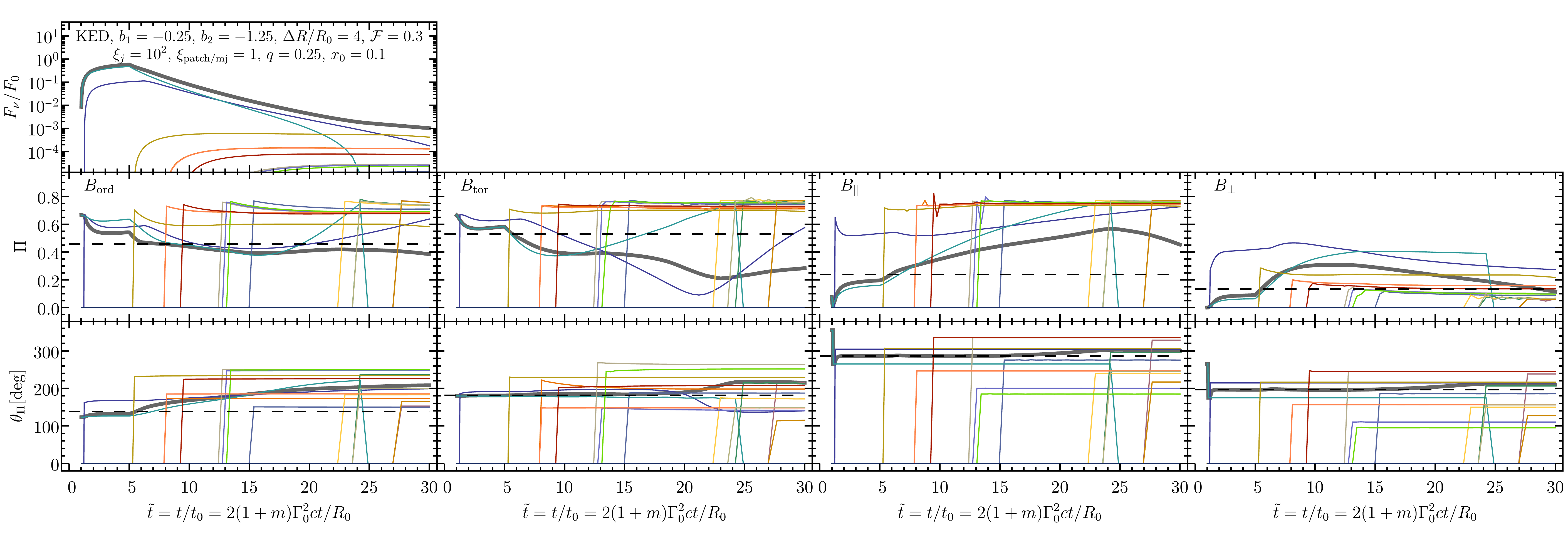}
    \caption{Same as Fig.\,\ref{fig:THmj-in-THJ-diff-Bfield} but with $\Delta R/R_0=4$.}
    \label{fig:THmj-in-THJ-diff-Bfield-dRoR-4}
\end{figure*}

\begin{figure*}
    \centering
    \includegraphics[width=0.23\textwidth]{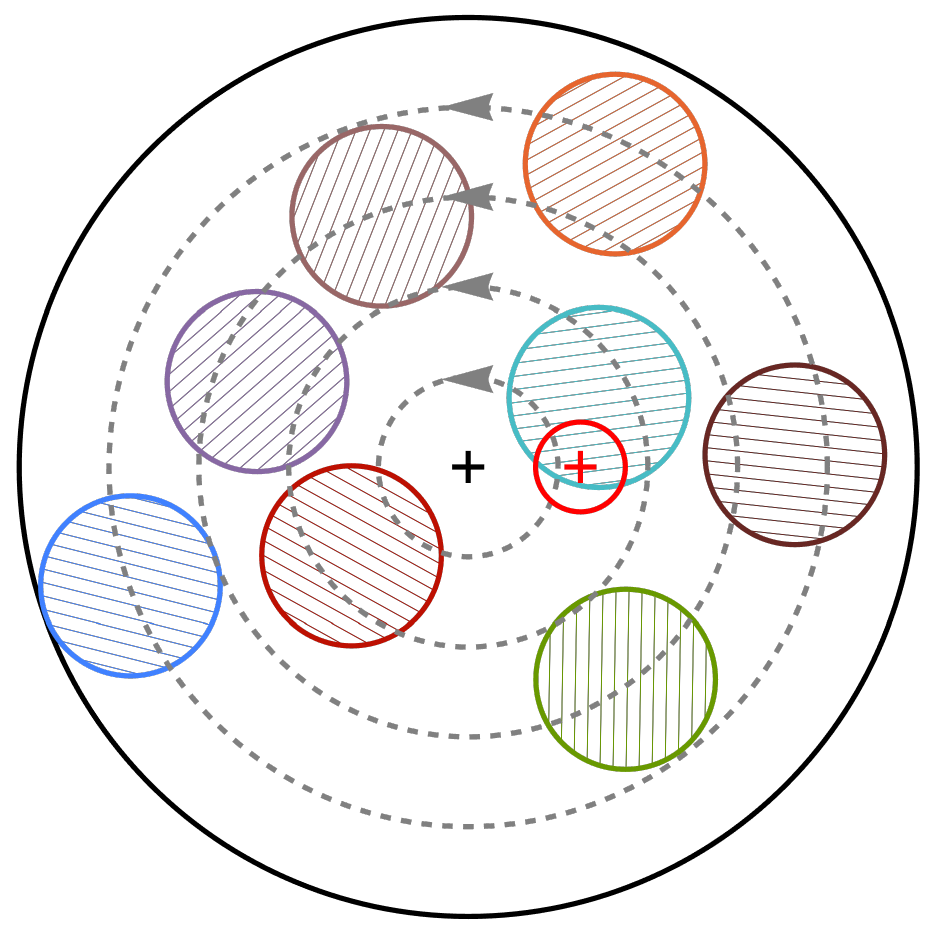} \\
    \centering
    \includegraphics[width=\textwidth]{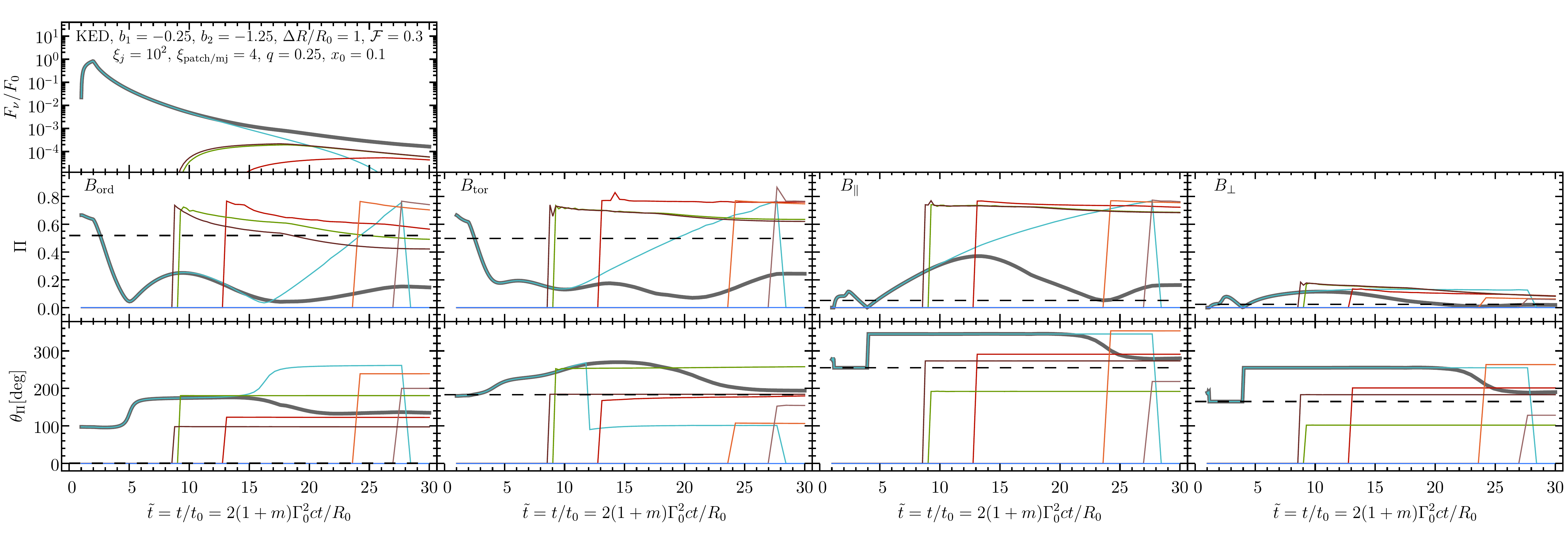}
    \caption{Same as Fig.\,\ref{fig:THmj-in-THJ-diff-Bfield} but with $\xi_{\rm patch/mj}=4$, 
    $N_{\rm patch/mj}=8$ and $\mathcal{F}=0.32$. }
    \label{fig:THmj-in-THJ-diff-Bfield-ximj-4}
\end{figure*}

\begin{figure*}
    \centering
    \includegraphics[width=0.23\textwidth]{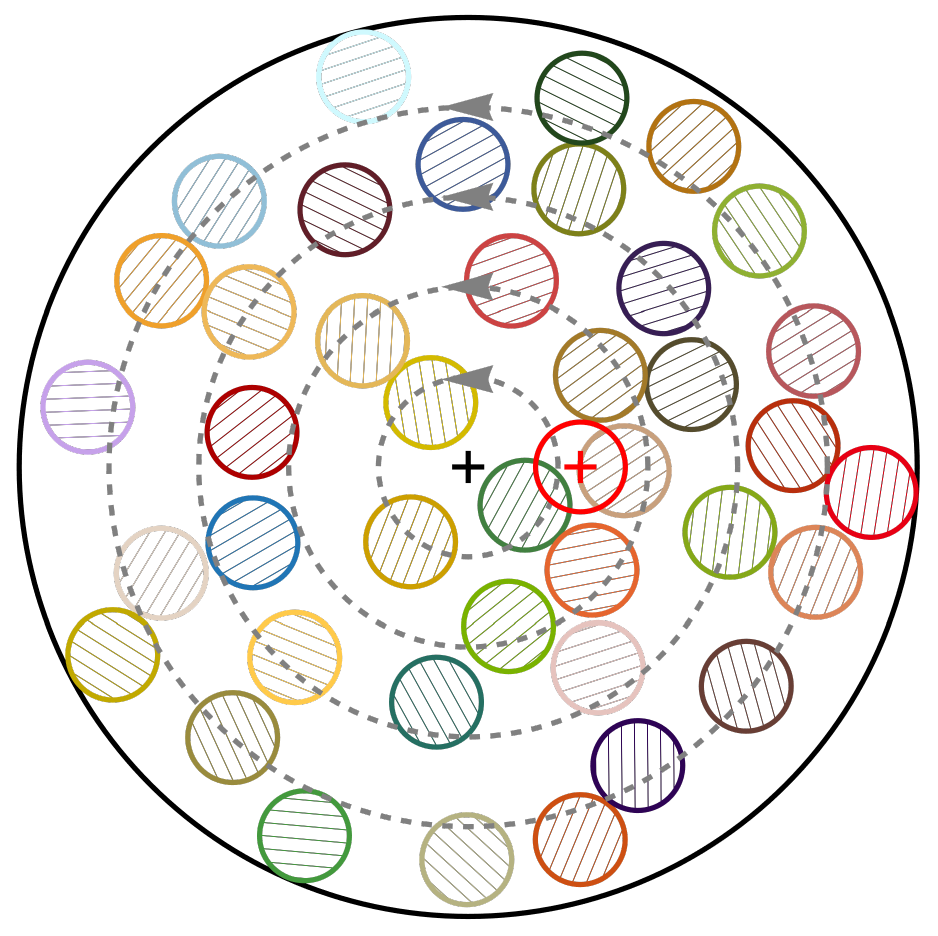} \\
    \centering
    \includegraphics[width=\textwidth]{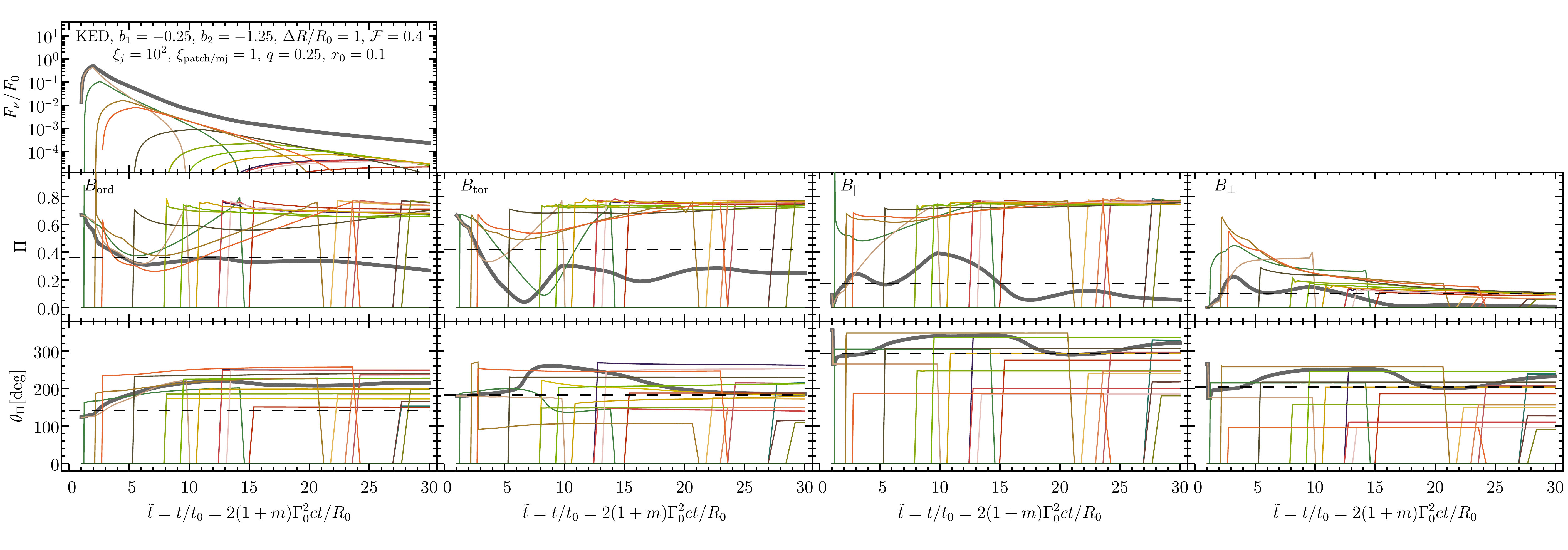}
    \caption{Same as Fig.\,\ref{fig:THmj-in-THJ-diff-Bfield} but with covering factor $\mathcal{F}=0.4$ 
    corresponding to $N_{\rm patch/mj}=40$. }
    \label{fig:THmj-in-THJ-diff-Bfield-Fcover-0.4}
\end{figure*}

\section{Non-Axisymmetric Jet Structure: Mini-Jets and Patches}\label{sec:non-axisymmetric-jets}

Here we develop a general formalism that describes the distribution of MJs/patches inside the aperture 
of the global jet. For brevity we only refer to MJs in all of the formulae in what follows, but the same formalism 
is used to describe patches.

We consider several (non-overlapping) MJs distributed inside a global collimated outflow (or jet), 
which is axisymmetric on average, with angular size much larger than each MJ. The coordinates 
$(\theta_{\rm mj}, \varphi_{\rm mj})$ of the symmetry axis of each MJ (relative to the symmetry axis 
of the global flow), as shown in Fig.\,\ref{fig:geometry}, are randomly drawn from their distribution 
according to the solid angle, $d\Omega=d(\cos\theta)d\varphi$, so that the probability density function for a given 
$0\leq\theta_{\rm mj}\leq(\theta_{\max}-\bar\theta_{\rm mj})$ and $0\leq\varphi_{\rm mj}\leq2\pi$ follows from
\begin{eqnarray}
    \label{eq:thmj-dist}
    P(\theta_{\rm mj})d\theta_{\rm mj} &=& \frac{\sin\theta_{\rm mj}d\theta_{\rm mj}}{[1-\cos(\theta_{\max}-\bar\theta_{\rm mj})]}\ , \\
    \label{eq:phimj-dist}
    P(\varphi_{\rm mj})d\varphi_{\rm mj} &=& \frac{d\varphi_{\rm mj}}{2\pi}\ .
\end{eqnarray}
Here $\bar\theta_{\rm mj}$ is the MJ half-opening angle 
and $\theta_{\max}$ can be the half-opening angle of a top-hat jet ($\theta_j$) or some multiple of the core angle 
($\theta_c$) of an angular structured jet. To obtain non-overlapping MJs, we set the condition so that the angular separation 
between the symmetry axes of any two neighbouring MJs follows $\Delta\theta^{\rm mj}_{i,j}\geq\bar\theta_{{\rm mj},i}+\bar\theta_{{\rm mj},j}$, 
or $\hat{n}_i\cdot\hat{n}_j\leq\cos(\bar\theta_{{\rm mj},i}+\bar\theta_{{\rm mj},j})$ 
where $\hat{n}_k$ is the 
direction of the $k^\textrm{th}$ MJ's symmetry axis, i.e. 
\begin{eqnarray}
\hat{n}_i\cdot\hat{n}_j=\cos\Delta\theta^{\rm mj}_{i,j} = &
\\
&\hspace{-2.05cm}\sin\theta_{{\rm mj},i}\sin\theta_{{\rm mj},j}\cos(\varphi_{{\rm mj},i}\!-\!\varphi_{{\rm mj},j})+ 
\cos\theta_{{\rm mj},i}\cos\theta_{{\rm mj},j}\,. \nonumber
\end{eqnarray}
While the minimum separation here is strictly true for non-overlapping top-hat MJs, it must be some multiple of the core sizes if the 
MJs have angular structure. The total number $N_{\rm mj}$ of MJs is limited by the covering factor, which e.g. in the 
case of top-hat MJs inside a top-hat global jet is given by 
\begin{equation}
    \mathcal{F} = N_{\rm mj}\fracb{\bar\theta_{\rm mj}}{\theta_j}^2 = N_{\rm mj}\frac{\xi_{\rm mj}}{\xi_j}\,,
\end{equation}
where $\xi_{\{\rm j,mj\}}\equiv(\Gamma\{\theta_j,\bar\theta_{\rm mj}\})^2$.

To calculate the pulse-profile and polarization from each MJ, we use the formalism described in \S\ref{sec:theory-model} 
to obtain the Stokes parameters $\{I_{\nu,\rm mj},Q_{\nu,\rm mj},U_{\nu,\rm mj}\}$. Since for incoherent radiation 
the Stokes parameters are additive, we then obtain the net lightcurve 
and polarization by adding the contribution from each MJ, such that $\Psi_{\nu}$ = $\sum_i \Psi_{\nu,{\rm mj},i}$ 
where $\Psi_\nu=\{I_\nu,Q_\nu,U_\nu\}$. The net polarization at time $t_z$ is obtained from 
$\Pi_\nu(t_z)=\sqrt{Q_\nu^2(t_z)+U_\nu^2(t_z)}\big/I_\nu(t_z)$ and polarization angle from 
$\theta_{\Pi}(t_z) = \frac{1}{2}\arctan[U_\nu(t_z)/Q_\nu(t_z)]$. The time-integrated polarization is likewise obtained 
from time-integrated Stokes parameters, with $\bar\Psi_\nu\,=\int\Psi_\nu(t_z)dt_z$ and $\bar\Pi_\nu=\sqrt{\bar Q_\nu^2 + \bar U_\nu^2}/\bar I_\nu$.

Depending on the scenario considered, the MJs/patches can vary in emissivity, 
bulk-$\Gamma$, angular size, and even angular structure (only in the case of MJs). Here we consider three different scenarios, 
namely:\vspace{0.1cm}\\ \vspace{0.1cm}\noindent
(\textbf{i}) top-hat MJs or uniform patches inside a global top-hat jet,\\ \vspace{0.1cm}\noindent
(\textbf{ii}) the same inside a global 
jet with angular structure, and\\ \vspace{0.1cm}\noindent
(\textbf{iii}) MJs with angular structure in emissivity in a global top-hat jet.

\subsection{Top-Hat \MJs~Inside a Global Top-Hat Jet}
The top-hat \MJs~have angular size (half-opening angle) $\bar\theta_{\rm mj} \Leftrightarrow \xi_{\rm mj}\equiv(\Gamma\bar\theta_{\rm mj})^2$ 
that is much smaller than that of the global top-hat jet with $\theta_j\Leftrightarrow\xi_j\equiv(\Gamma\theta_j)^2$, where 
$\Gamma\gg1$ is the bulk LF of the entire flow. The distribution of the \MJs~inside the global jet is shown  
in Fig.\,\ref{fig:THmj-in-THJ-diff-Bfield}, along with the location of the viewing angle surrounded by the beaming cone (red circle).
Fig.\,\ref{fig:THmj-in-THJ-diff-Bfield} also shows the B-field lines for the globally ordered toroidal field ($B_{\rm tor}$) as well 
as the orientation of the B-field lines for the case of an ordered field ($B_{\rm ord}$) inside of every patch.

The normalized lightcurve, net polarization and polarization angle, are shown as a function of normalized time 
$\tilde t=t/t_0=t_z/t_{0,z}$ in Fig.\,\ref{fig:THmj-in-THJ-diff-Bfield} for all of the different B-field configurations 
and for a pulse with $\Delta R/R_0=1$ and $x_0=0.1$ for all of the \MJs. Contributions from different \MJs~are shown with 
different colors that correspond to their locations within the global jet. The lightcurve is normalized using the maximum 
flux density $F_0$ from a hypothetical (if not present) MJ/patch centered at the LOS. Since all of the \MJs~are viewed 
off-axis (LOS outside of the MJ/patch aperture) in this setup, 
the emission arrives at the observer with an angular delay in addition to the radial one (that causes the emission 
from a MJ/patch along the LOS to start arriving at $\tilde{t}=1$). Hence, the presence of an 
offset in the rise time of the pulse with respect to $t=t_0$ ($\tilde{t}=1$). For reference, the angular delay time for emission arriving 
from an angular distance of $1/\Gamma$ (size of beaming cone) away from the LOS is exactly the same as the radial time 
delay for a uniform flow (with $m=0$), which would give a total time delay of $t=2t_0$ or $\tilde t = 2$. The lightcurve also 
shows a shallower decay post-peak, in comparison to that of individual \MJs, due to the contribution from several \MJs~with 
progressively larger offsets in the rise times. 

The polarization curves, as shown in the middle panels of Fig.\,\ref{fig:THmj-in-THJ-diff-Bfield}, 
initially reflect the $\Pi$ of the MJ/patch that dominates the flux, which in this case is the one closest to the LOS. 
As emission from other \MJs~starts to arrive, and even dominate the flux at different times, the net $\Pi$ declines much 
more rapidly in comparison to the very shallow decay of that from each MJ/patch. The reason behind this decline in the net 
$\Pi$ is the cancellation due to different randomly oriented PAs of the overlapping emission, which is also reflected in the temporal evolution 
of the net PA ($\theta_\Pi$) as shown in the bottom panels of Fig.\,\ref{fig:THmj-in-THJ-diff-Bfield}. The net polarization 
is then approximately obtained from $\Pi\sim\Pi_0/\sqrt{N}$, where $\Pi_0$ is the polarization 
at any given time from a MJ/patch that makes the dominant contribution to the flux. This polarization is then diluted by the 
contribution from $N-1$ other \MJs~that effectively contribute to the total flux. Therefore, as $N$ grows the net polarization will decline 
accordingly. The characteristic trend of net $\Pi$ in this scenario is that it will be largest, 
even close to maximal, at the onset of the peak, but it will always decline in the tail of the pulse while the net PA 
changes continuously. 

The time-integrated polarization is shown using a black dashed line in Fig.\,\ref{fig:THmj-in-THJ-diff-Bfield}. When $q<1$ 
in a uniform jet, the time-integrated polarization vanishes for B-fields, e.g. $B_\perp$ and $B_\parallel$, that are axisymmetric 
around the local shock normal or velocity direction (radial here), due to complete cancellation in the Stokes plane. Here, the MJs break that 
symmetry and yield a modestly high polarization.

Although the global PA changes continuously, that of each MJ is still strictly limited to $\Delta\theta_\Pi=90^\circ$ 
in the cases of $B_\perp$ and $B_\parallel$. This is due to the fact that each MJ is axisymmetric around its own symmetry axis.   
When the PA changes by $90^\circ$, the polarization vanishes and reappears. The temporal evolution of PA is exactly the same 
for both $B_\perp$ and $B_\parallel$ fields, except for the $90^\circ$ offset. Since the field structure is similar in both cases, 
i.e. axisymmetric around the radial direction (generally it is around the local shock normal), the temporal evolution of the PA is 
also similar, and the 90$^\circ$ offset is caused by the same offset in the orientations of the two B-fields with respect to the 
radial direction. 

The axisymmetry is broken in both the $B_{\rm ord}$ and $B_{\rm tor}$ cases due to the presence of a large-scale B-field, and 
in both the PA does change continuously and gradually. In some instances, when the global $\Pi$, and even that from a single patch, shows a 
local minimum with $\Pi$ close to zero, the PA correspondingly shows a gradual change with $\Delta\theta_\Pi<90^\circ$, akin to an 
S-curve and not a sharp one.

Figure\,\ref{fig:THmj-in-THJ-diff-Bfield} only shows one random realization of the distribution of the \MJs. To see how the polarization 
properties might change with different arrangements, we produce two additional random realizations in Fig.\,\ref{fig:THmj-in-THJ-diff-Bfield-diff-seed} 
for comparison. Here we demonstrate how the temporal evolution of polarization can be significantly different in different 
bursts under the assumed scenario. More importantly, the time-integrated polarization can be drastically different. This is especially 
true for $B_\perp$, $B_\parallel$, and $B_{\rm ord}$ magnetic fields, where the time-integrated polarization can vary by $15-20$ per cent.

The effect of a longer emission time, with $\Delta R/R_0=4$, is shown in Fig.\,\ref{fig:THmj-in-THJ-diff-Bfield-dRoR-4}. This 
results in stretching the time when the last photon is received from the material closest to the LOS by a factor of 
$\hat R_f = 1+\Delta R/R_0$ for a (coasting) flow with $m=0$ \citep[see][for different critical times and additional details]{Gill-Granot-21}. 
Therefore, compared to the case shown in Fig.\,\ref{fig:THmj-in-THJ-diff-Bfield} with $\hat R_f=2$, the first break in the 
lightcurve shown in Fig.\,\ref{fig:THmj-in-THJ-diff-Bfield-dRoR-4}, corresponding to emission from $\hat R_f=5$, occurs 
at a normalized time that is 5/2 times larger. For some \MJs, the first break in the lightcurve may occur due 
to passage of the spectral break across the observed frequency $\nu$ as set by $x_0=\nu/\nu_0$. 

Figure\,\ref{fig:THmj-in-THJ-diff-Bfield-ximj-4} shows the effect of a larger $\xi_{\rm patch/mj}$. 
One obvious outcome for a fixed covering factor is that the total number of \MJs~must be smaller. Second, the 
post-lightcurve-peak contribution, arising from angles outside of the beaming cone, to the total emission from each 
MJ/patch is stretched out over larger times. As a result, changes in $\Pi$ and $\theta_{\Pi}$ show a temporal delay. 
Figure\,\ref{fig:THmj-in-THJ-diff-Bfield-diff-ximj} shows how the pulse profiles and polarization properties change 
for a variety of MJ/patch angular sizes. Inhomogeneities with angular sizes somewhat smaller than the local causal 
size, with $\xi_{\rm patch/mj}<1$, may potentially exist over a single dynamical time or so before being smeared 
out over the local causal size. Therefore, for comparison, Fig.\,\ref{fig:THmj-in-THJ-diff-Bfield-diff-ximj} 
also shows a case with $\xi_{\rm patch/mj}=1/4$.

Figure\,\ref{fig:THmj-in-THJ-diff-Bfield-Fcover-0.4} shows the effect of a larger covering factor with $\mathcal{F}=0.4$. 
A larger covering factor while keeping the angular size of the MJs/patches fixed means that a larger number of MJs/patches contribute 
to the global lightcurve at any given time, thus making the lightcurve somewhat smoother and the global polarization 
slightly smaller when compared with the $\mathcal{F}=0.3$ case. 
Different covering factors for a fixed $\xi_{\rm patch/mj}$ are shown in Fig.\,\ref{fig:THmj-in-THJ-diff-Bfield-diff-Fcover}. In all 
cases, the initial trend at $\tilde t \lesssim 10$ is the same regardless of $\mathcal{F}$ and the behavior only diverges at later 
times. This is due to the fact that the arrangement of the \MJs~within an angular size of a few beaming cones is the same in 
all cases. Being closer to the LOS these \MJs~make the dominant contribution at early times (or even the only contribution due 
to the larger angular time delay in the onset of the emission from more distant \MJs). An increase in the covering factor 
introduces additional \MJs, which, being farther away from the LOS, contribute to the total flux at later times. Hence the divergence 
in the polarization trends at later times. The scenario with the larger $\mathcal{F}$, and therefore larger $N_{\rm patch/mj}$, shows the 
lowest level of polarization at late times. This again confirms the aforementioned point where observation of larger number 
of \MJs~with randomly oriented PAs reduce the global $\Pi$.

\begin{figure*}
    \centering
    \includegraphics[width=0.23\textwidth]{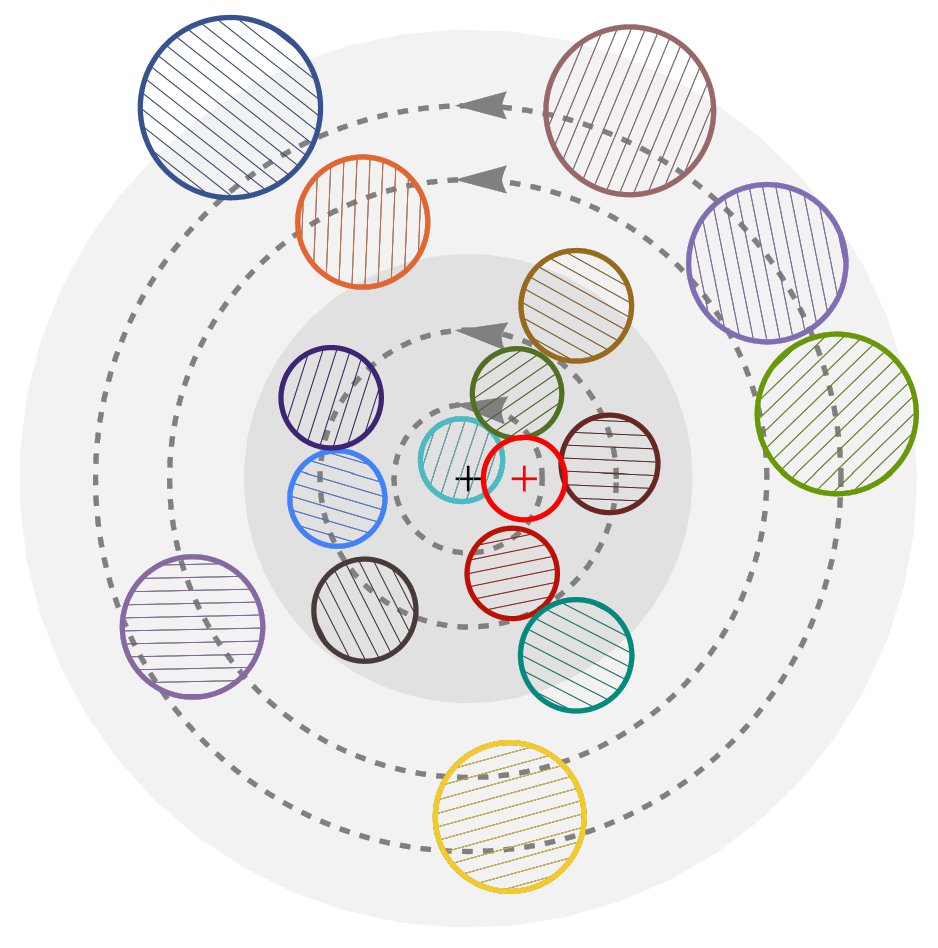} \\
    \includegraphics[width=\textwidth]{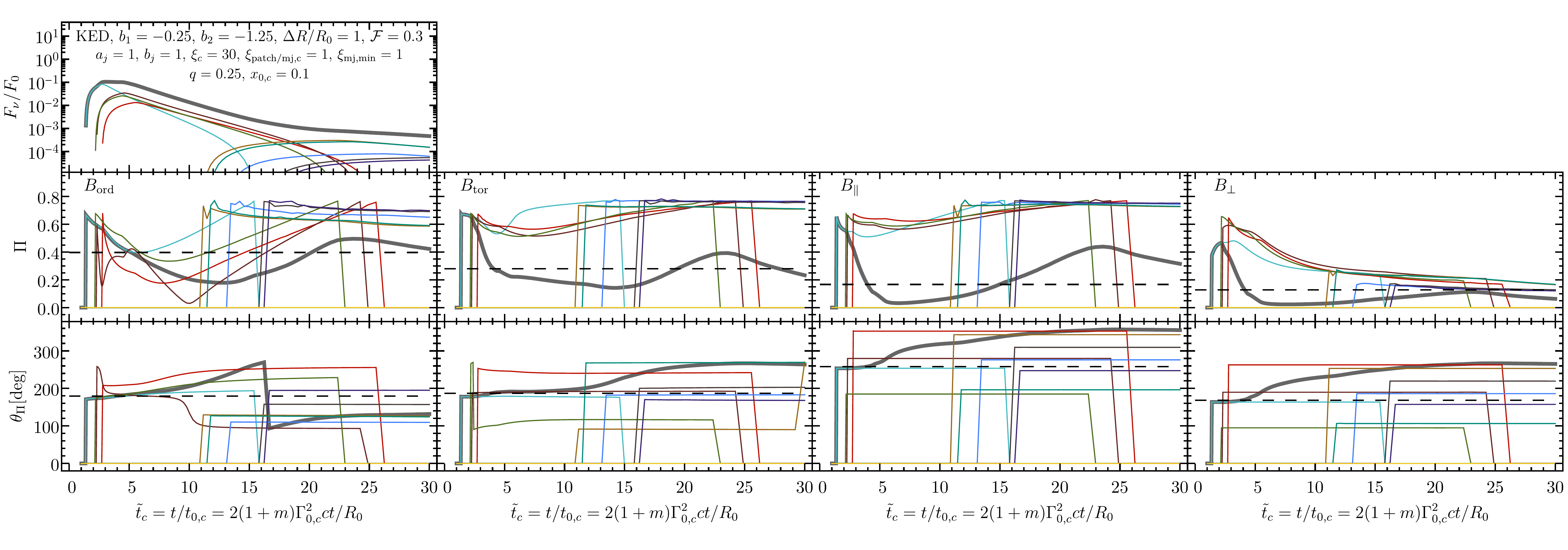}
    \caption{
    {\bf Top:} Top-hat \MJs~within the aperture of a global jet with power-law angular structure in both the 
    comoving emissivity ($a_j=1$) and bulk-$\Gamma$ ($b_j=1$). The global jet has a core with normalized angular size 
    $\xi_c^{1/2}=\Gamma_c\theta_c=\sqrt{30}$. The angular sizes of the \MJs~are assumed to scale according to 
    Eq.~(\ref{eq:thbar_mj}) with $\xi_{\rm patch/mj,c}=1$ and a covering factor $\mathcal{F}=0.3$. 
    Here, the red circle represents the $1/\Gamma(\theta_{\rm mj})$ beaming cone around the LOS (shown with a plus 
    symbol) located at $q \equiv \theta_{\rm obs}/\theta_c = 0.25$. The angular 
    sizes of the inner and outer gray shaded disks are $\theta_c$ and 2$\theta_c$. 
    {\bf Bottom:} Pulse profile and polarization evolution for different B-field configurations. 
    See Fig.\,\ref{fig:THmj-in-THJ-diff-Bfield} and text for more details.
    }
    \label{fig:THmj-in-PL-jet}
\end{figure*}

\begin{figure*}
    \centering
    \includegraphics[width=0.23\textwidth]{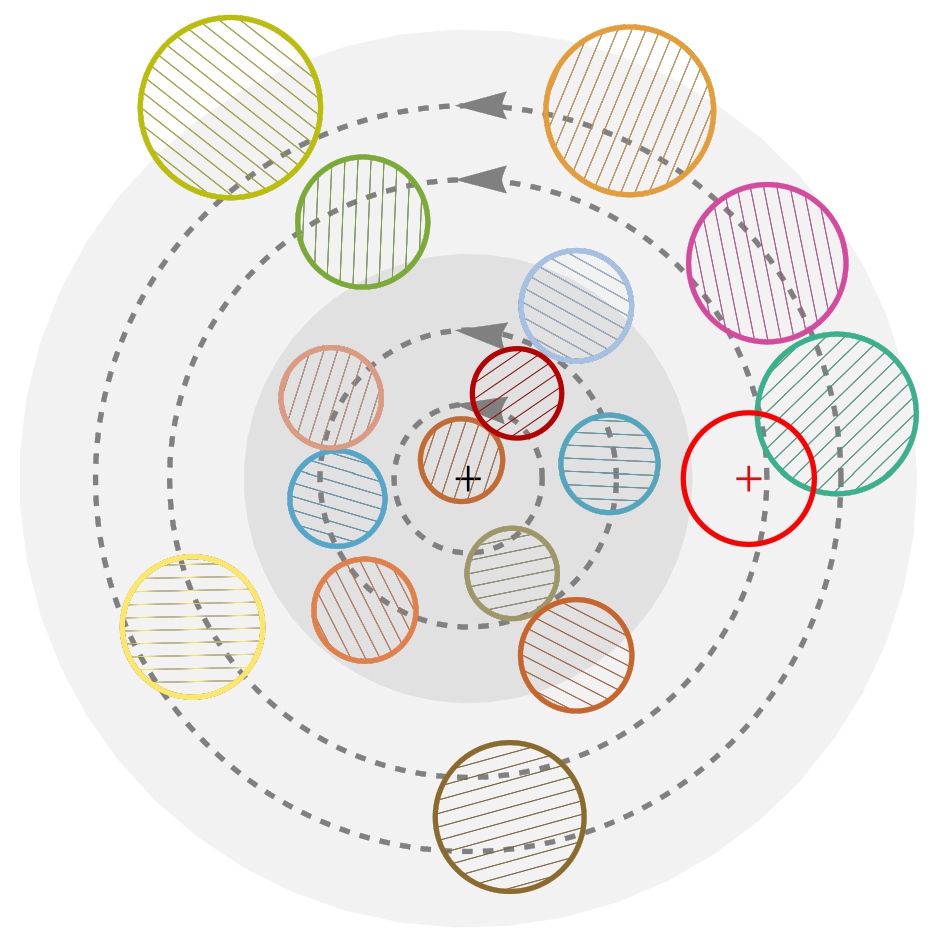} \\
    \includegraphics[width=\textwidth]{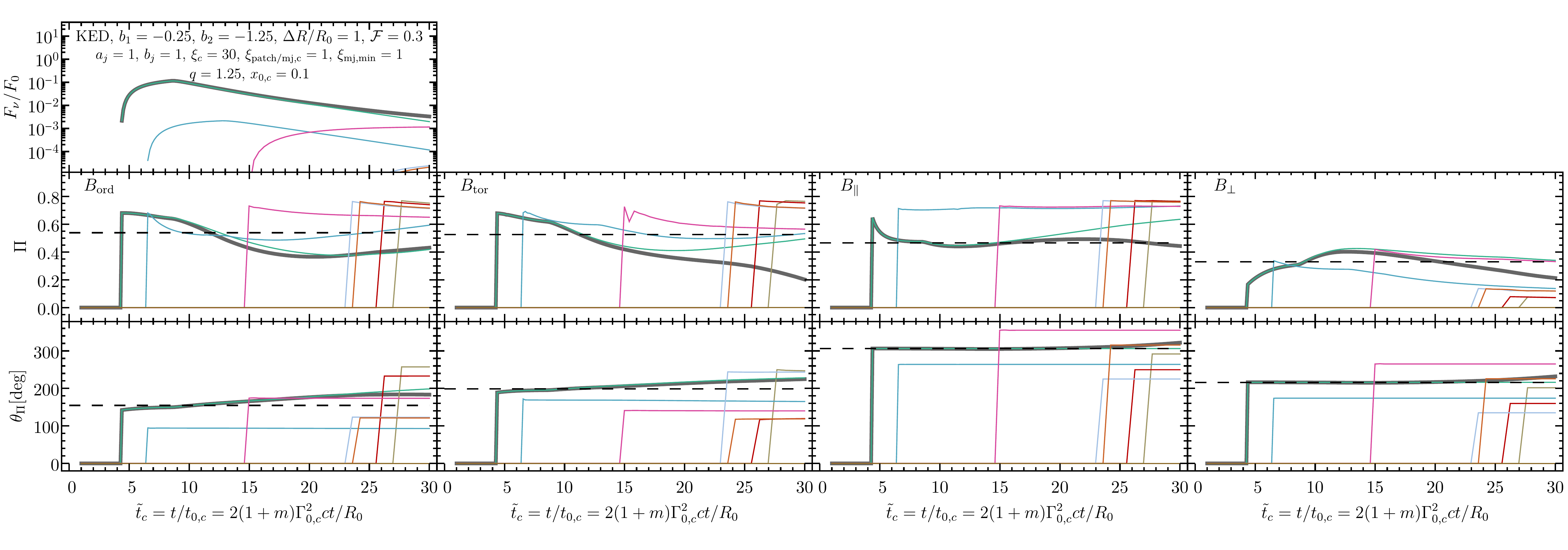}
    \caption{Same as Fig.\,\ref{fig:THmj-in-PL-jet} but with $q=1.25$.
    }
    \label{fig:THmj-in-PL-jet-q-1.25}
\end{figure*}

\begin{figure*}
    \centering
    \includegraphics[width=0.23\textwidth]{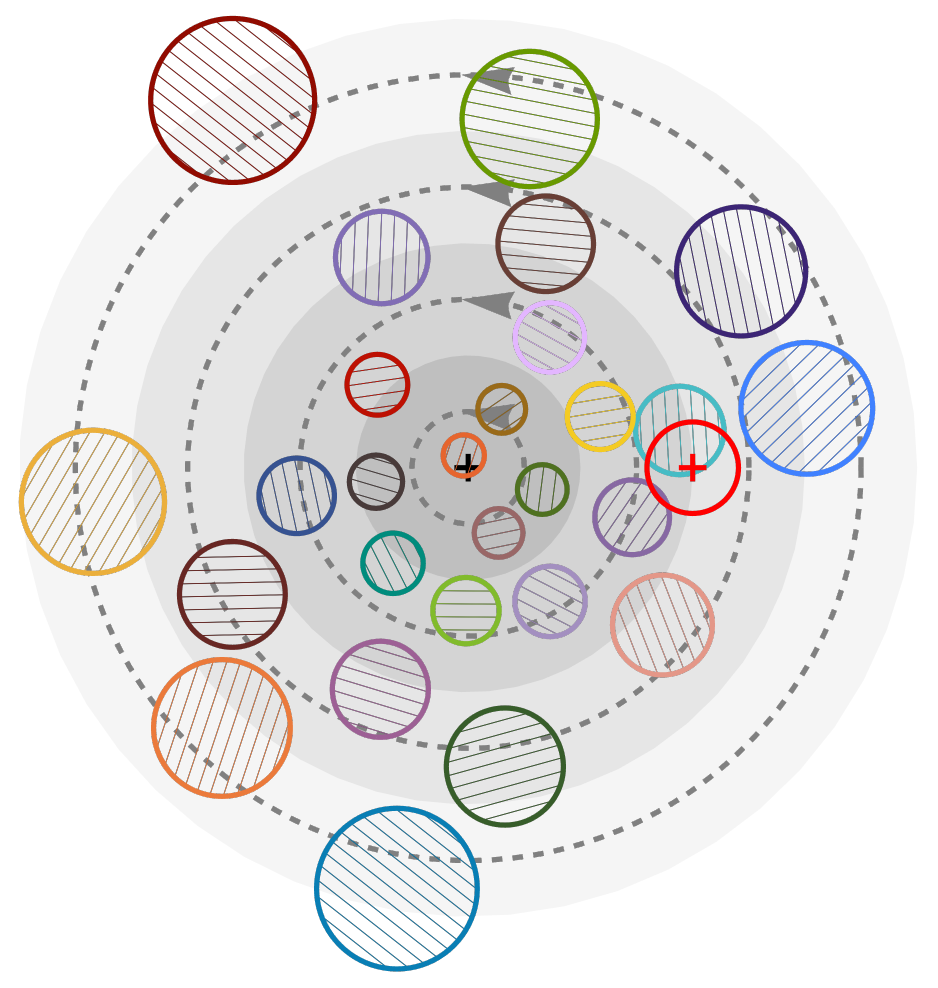} \\
    \includegraphics[width=\textwidth]{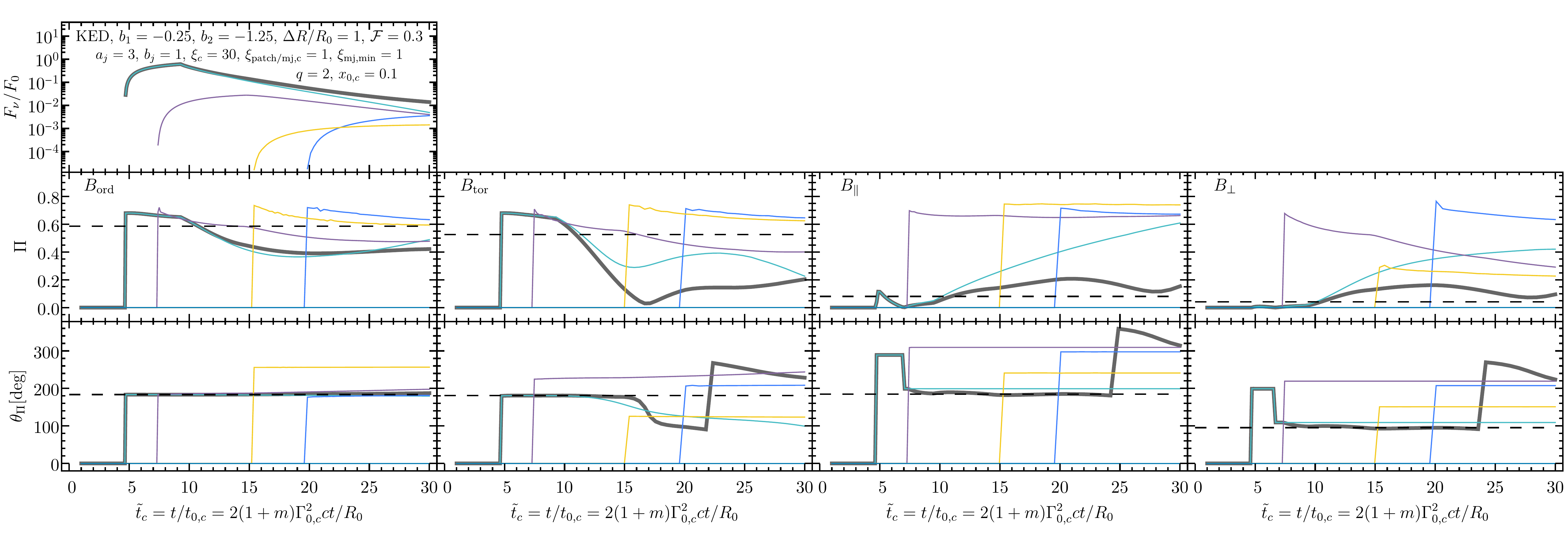}
    \caption{Same as Fig.\,\ref{fig:THmj-in-PL-jet} but with $a_j=3$ and $q=2$. The four gray shaded disks 
    have angular sizes $\theta_c$, $2\theta_c$, $3\theta_c$, $4\theta_c$.}
    \label{fig:THmj-in-PL-jet-a-3-q-2}
\end{figure*}

\begin{figure*}
    \centering
    \includegraphics[width=0.23\textwidth]{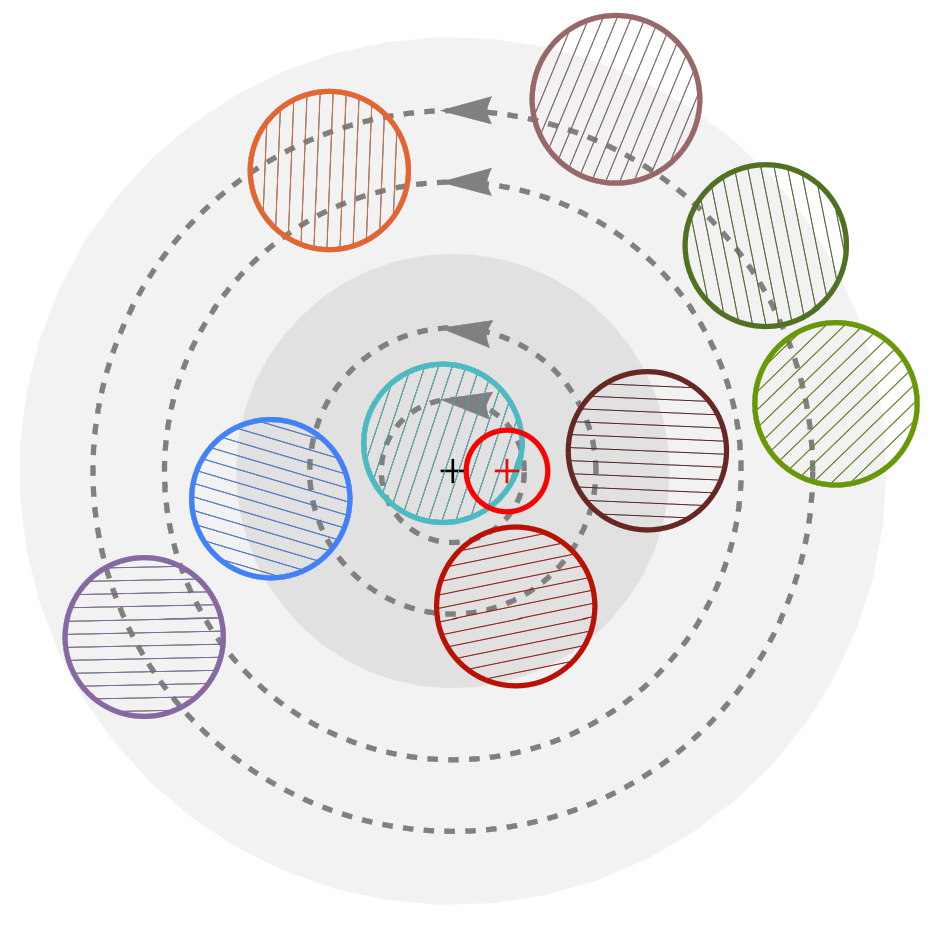} \\
    \includegraphics[width=\textwidth]{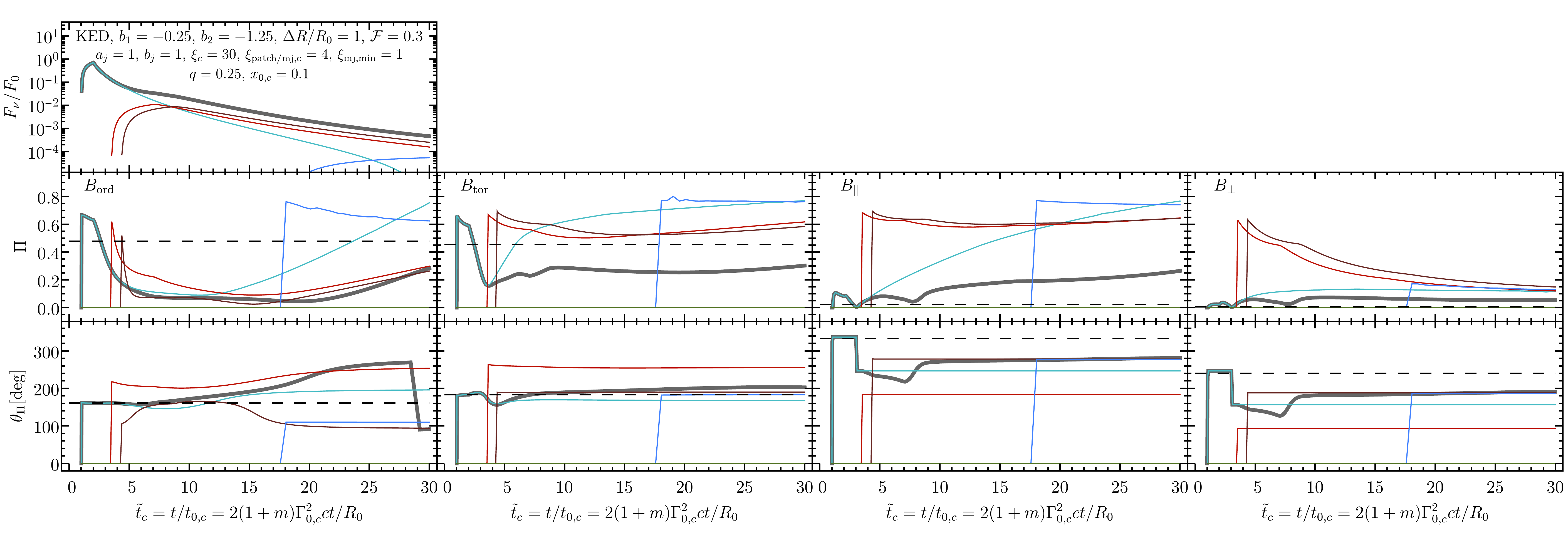}
    \caption{Same as Fig.\,\ref{fig:THmj-in-PL-jet} but with $\xi_{\rm mj,c}=4$.}
    \label{fig:THmj-in-PL-jet-ximj-4}
\end{figure*}

\begin{figure*}
    \centering
    \includegraphics[width=0.23\textwidth]{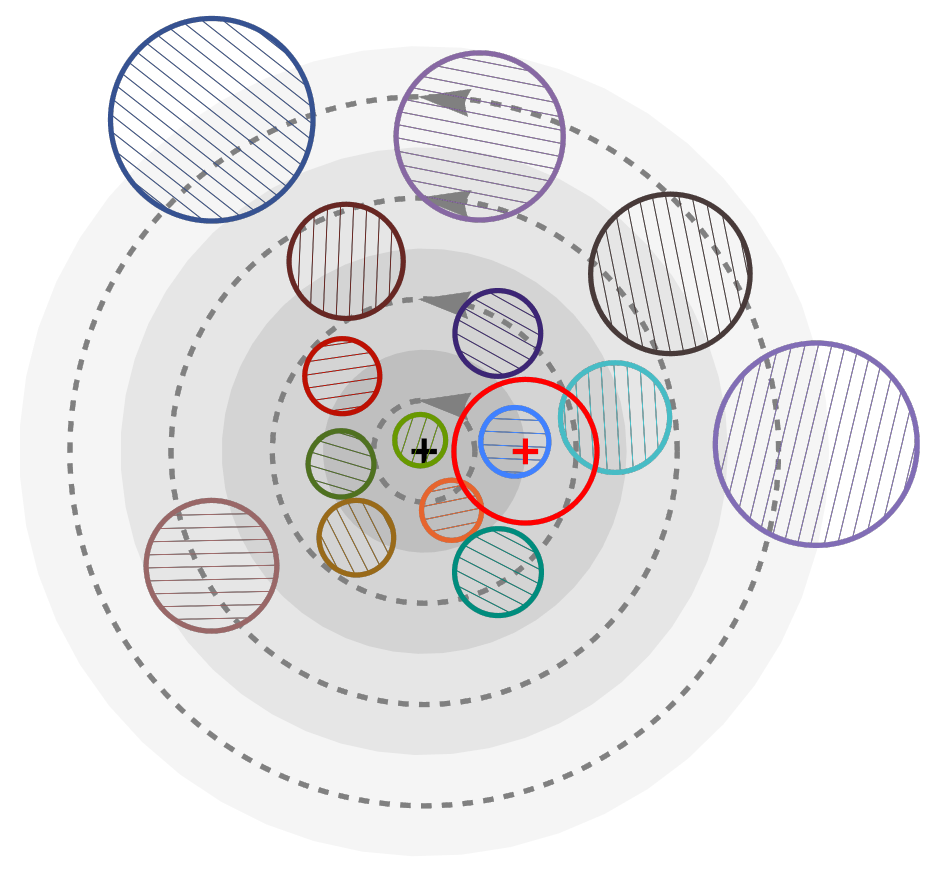} \\
    \includegraphics[width=\textwidth]{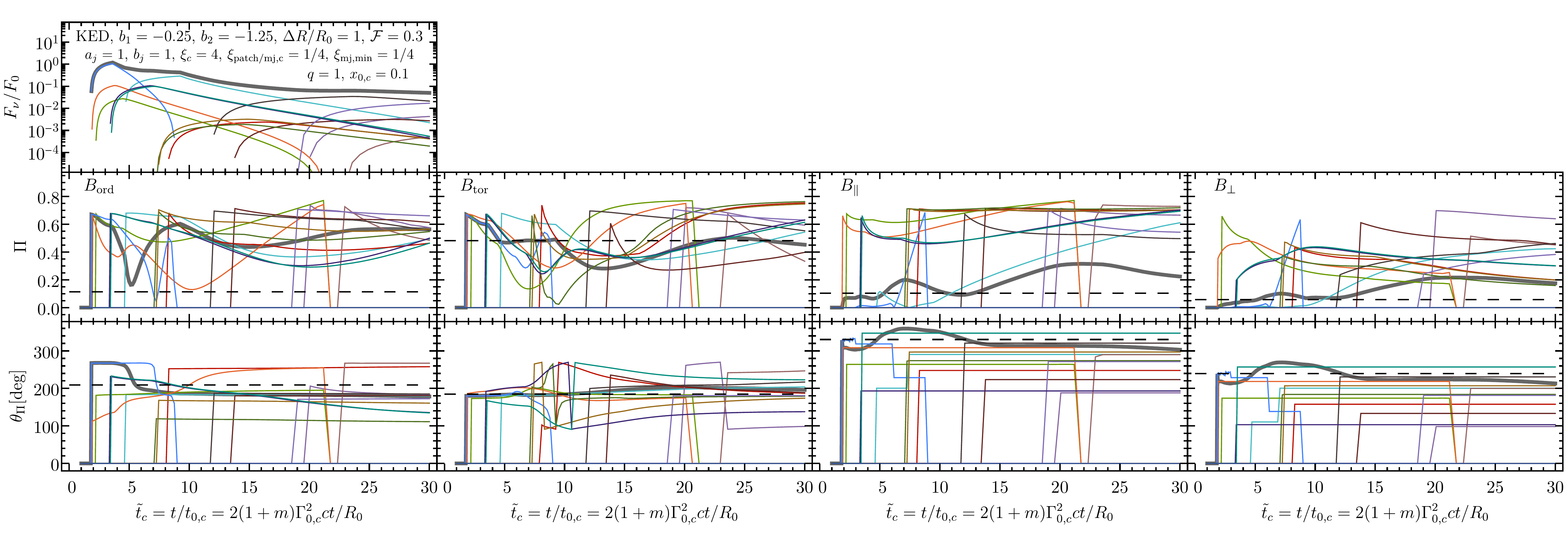}
    \caption{Same as Fig.\,\ref{fig:THmj-in-PL-jet}, but with $\xi_c=4$, $\xi_{\rm mj,c}=\xi_{\rm mj,\min}=1/4$, 
    $q=1$, and $\mathcal{F}=0.3$. The four gray shaded disks have angular sizes $\theta_c$, $2\theta_c$, $3\theta_c$, $4\theta_c$.}
    \label{fig:THmj-in-PL-jet-xic-4-q-1-F-0.3}
\end{figure*}

\subsection{Top-Hat \MJs~in an Angular Structured Global Jet}
Next, we consider a global jet with angular structure both in the comoving spectral peak luminosity and bulk-$\Gamma$ 
\citep[e.g.,][]{Rossi+02,Granot-Kumar-03,Kumar-Granot-03,Rossi+04,Gill-Granot-18a,Gill+20},
\begin{eqnarray}\label{eq:ang-profile}
\frac{\nu_p'L'_{\nu_p'}(\theta_{\rm mj})}{L'_0} = \Theta^{-a_j},\quad\quad\quad\quad
\frac{\Gamma(\theta_{\rm mj})-1}{\Gamma_c-1} = \Theta^{-b_j},
\\ \nonumber
\Theta(\theta_{\rm mj}) = \sqrt{1+\fracb{\theta_{\rm mj}}{\theta_c}^2}\,,\quad\quad\quad\quad\quad\quad   
\end{eqnarray}
where $\nu_p'$ is the spectral peak frequency, $L_0' = \nu_p'L_{\nu_p'}'(\theta_{\rm mj}=0)$ is the spectral peak 
luminosity at the global jet symmetry axis, $\theta_{\rm mj}$ is the polar angle of the MJ/patch symmetry axis, and the angular profiles fall off as a power law for 
polar angles larger than some core angle $\theta_c$. Since $\bar\theta_{\rm mj}\ll\theta_{\max}\equiv\kappa\theta_c$, 
where $\kappa\sim$\,few is limited by the compactness argument \citep{Gill+20}, we consider only top-hat \MJs~but whose $L'_{\nu'}$ and/or 
bulk-$\Gamma$ vary with $\theta_{\rm mj}$ according to the angular structure profile. 

When $\Gamma_{\rm mj}=\Gamma(\theta_{\rm mj})$ of the \MJs~is varied in the sample, three important 
effects occur that are not present in the earlier case of a global top-hat jet. First, the angular size ($\bar\theta_{\rm mj}$) 
of the \MJs~will vary according to the measure $\xi_{\rm mj}=(\Gamma_{\rm mj}\bar\theta_{\rm mj})^2$, where for a fixed $\xi_{\rm mj}$ the 
angular size $\bar\theta_{\rm mj}\propto1/\Gamma_{\rm mj}$ will gradually grow as $\Gamma_{\rm mj}$ declines. 
The lowest physical value of $\xi_{\rm mj}$ is unity since that represents the smallest angular scale 
over which the jet remains causally connected, and any smaller-scale large amplitude inhomogeneities 
in the flow will naturally expand to this size over a dynamical time. To be able to see the polarization 
trends in a simpler manner, we instead assume a broken power-law approximation for the growth of $\bar\theta_{\rm mj}$ 
with $\theta_{\rm mj}$, such that
\begin{equation}\label{eq:thbar_mj}
\bar\theta_{\rm mj}(\theta_{\rm mj}) = \max\left[\bar\theta_{\rm mj,c},\frac{\sqrt{\xi_{\rm mj,\min}}}{\Gamma(\theta_{\rm mj})}\right]\,,\quad\ 
    \bar\theta_{\rm mj,c} = \frac{\sqrt{\xi_{\rm mj,c}}}{\Gamma_c} 
    =\theta_c\sqrt{\frac{\xi_{\rm mj,c}}{\xi_c}}\,,
\end{equation}
where $\xi_c=(\Gamma_c\theta_c)^2$, $\xi_{\rm mj,c}=(\Gamma_c\bar\theta_{\rm mj,c})^2$ and $\bar\theta_{\rm mj,c}=\bar\theta_{\rm mj}(\theta_{\rm mj}=0)$. 
When being more conservative $\xi_{\rm mj,\min}=1$, but here we also allow for the possibility of $\xi_{\rm mj,\min}\leq1$. 
In the ultrarelativistic limit, the above condition can be expressed as 
\begin{eqnarray}
    \xi_{\rm mj} &=& (\Gamma_{\rm mj}\bar\theta_{\rm mj})^2=\max\left[\xi_{\rm mj,c}\left(\frac{\Gamma_{\rm mj}}{\Gamma_c}\right)^2,\xi_{\rm mj,\min}\right] \\
    &\approx& \max\left[\xi_{\rm mj,c}\Theta^{-2b_j},\xi_{\rm mj,\min}\right]\,. \nonumber
\end{eqnarray}
It implies that the angular scale of the \MJs~remains constant for a given $\xi_{\rm mj,c}$ and $\Gamma_c\gg1$ or $\theta_c\ll1$, until it 
becomes smaller than $\sqrt{\xi_{\rm mj,\min}}/\Gamma_{\rm mj}$ after which point $\bar\theta_{\rm mj}=\sqrt{\xi_{\rm mj,\min}}/\Gamma_{\rm mj}$. 
In the ultrarelativistic limit, this critical angle is given by 
\begin{equation}
    \theta_{\rm mj\star} = \theta_c\left[\left(\frac{\xi_{\rm mj,c}}{\xi_{\rm mj,\min}}\right)^{1/b_j}-1\right]^{1/2}
\end{equation}
for $b_j>0$. 
In the limiting case of  $\xi_{\rm mj,c}=\xi_{\rm mj,\min}$, $\theta_{\rm mj\star}=0$ and the growth of $\bar\theta_{\rm mj}$ occurs 
for all angles $\theta_{\rm mj}$. To maintain a uniform covering factor locally when $\bar\theta_{\rm mj}$ grows with angle, 
the probability density in Eq.\,(\ref{eq:thmj-dist}) is modified to 
 \begin{equation}
     P(\theta_{\rm mj}>\!\theta_{\rm mj\star}) \propto \frac{\sin\theta_{\rm mj}}{\bar\theta_{\rm mj}^2} \sim \frac{\theta_{\rm mj}}{\bar\theta_{\rm mj}^2}\,,
 \end{equation}
where the approximate expression on the right is valid for $\theta_{\rm mj}\ll1$. Finally, since the MJs/patches have different 
bulk $\Gamma$, the radial distance $R_0$ travelled by them before they start radiating can be different. For simplicity, here 
we assume that all MJs/patches start to radiate at the same radius $R_0$.
 
The covering factor in this case is calculated from 
\begin{equation}
    \mathcal F = \sum_i\left[\frac{\bar\theta_{{\rm mj},i}(\theta_{\rm mj})}{\theta_{\rm max}}\right]^2\,,
\end{equation}
and a round number of \MJs~are obtained for a given $\mathcal{F}$. 
Second, assuming that the initial comoving spectral peak frequency ($\nu_0'$) at $R=R_0$ is the same throughout the flow, the observed 
frequency of these first photons received along the LOS of an on-axis observer for each MJ/patch will be different according to 
$x_{0,\rm mj}=\nu/\nu_{0,\rm mj}$, where $\nu_{0,\rm mj} = 2\Gamma_{0,\rm mj}\nu_0'/(1+z)$. For a fixed $\nu_0'$, this only 
depends on the angular profile of the global flow, such that 
$x_{0,\rm mj}(\theta_{\rm mj})=x_{0,c}[\Gamma_{0,c}/\Gamma_0(\theta_{\rm mj})]$. Third, the normalized arrival time of these  
first photons (accounting only for the radial time delay), 
$\tilde t_{\rm mj} = t/t_{0,\rm mj} = 2(1+m)\Gamma_{0,\rm mj}^2ct/R_0 = \tilde t_c[\Gamma_0(\theta_{\rm mj})/\Gamma_{0,c}]^2$, 
will be different for each MJ/patch. In practice, at most one MJ/patch intersects the LOS and the true arrival time of the first photon is larger by the corresponding angular time (and their Doppler factor is lower).

\begin{figure*}
    \centering
    \includegraphics[width=0.23\textwidth]{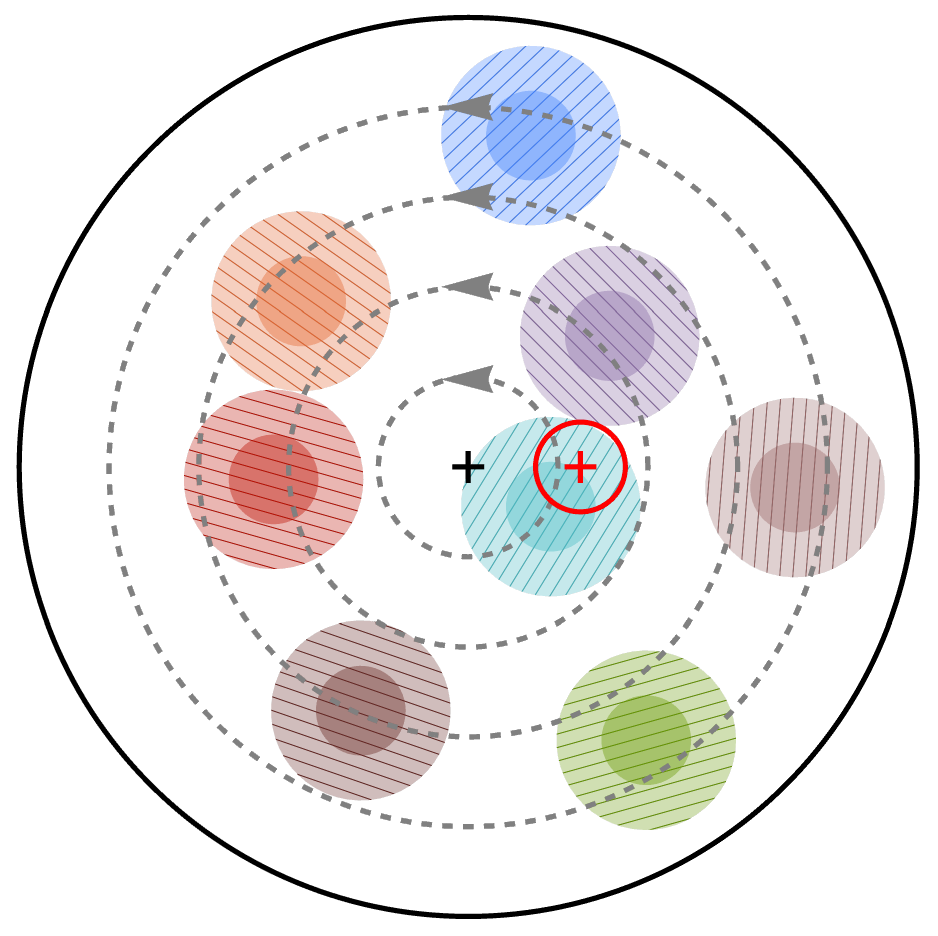} \\
    \includegraphics[width=\textwidth]{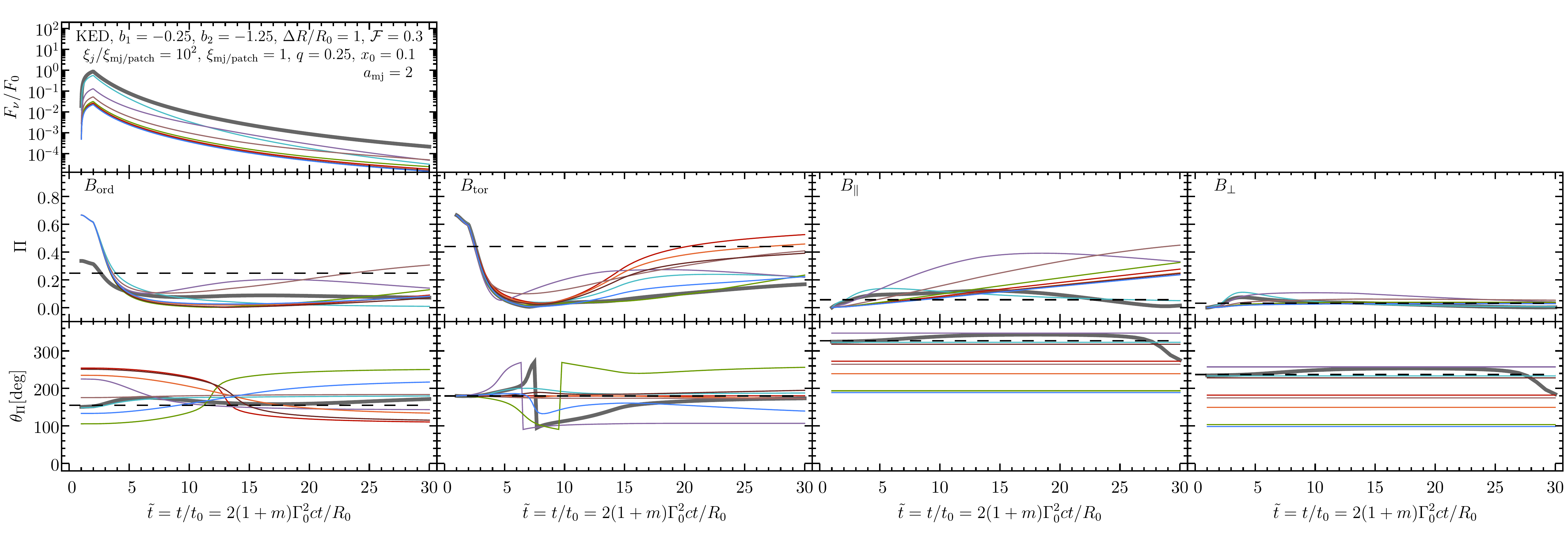}
    \caption{
    {\bf Top:} \MJs~with a power-law (comoving) emissivity profile, with power-law index $a_{\rm mj}=2$, 
    within the aperture of a global top-hat jet. The two shaded disks within each MJ/patch show the angular size 
   $\bar\theta_{\rm mj}=\sqrt{\xi_{\rm mj}}/\Gamma$ and $2\bar\theta_{\rm mj}$. The color gradient 
    reflects the gradient in the emissivity. The red circle shows the angular size of the $1/\Gamma$ beaming cone for comparison. 
    The lightcurve is normalized by the maximum flux density from a MJ/patch observed on-axis. 
    {\bf Bottom:} Pulse profile and polarization evolution shown for different B-field configurations. 
    }
    \label{fig:PLmj-in-TH-jet}
\end{figure*}

\begin{figure*}
    \centering
    \includegraphics[width=\textwidth]{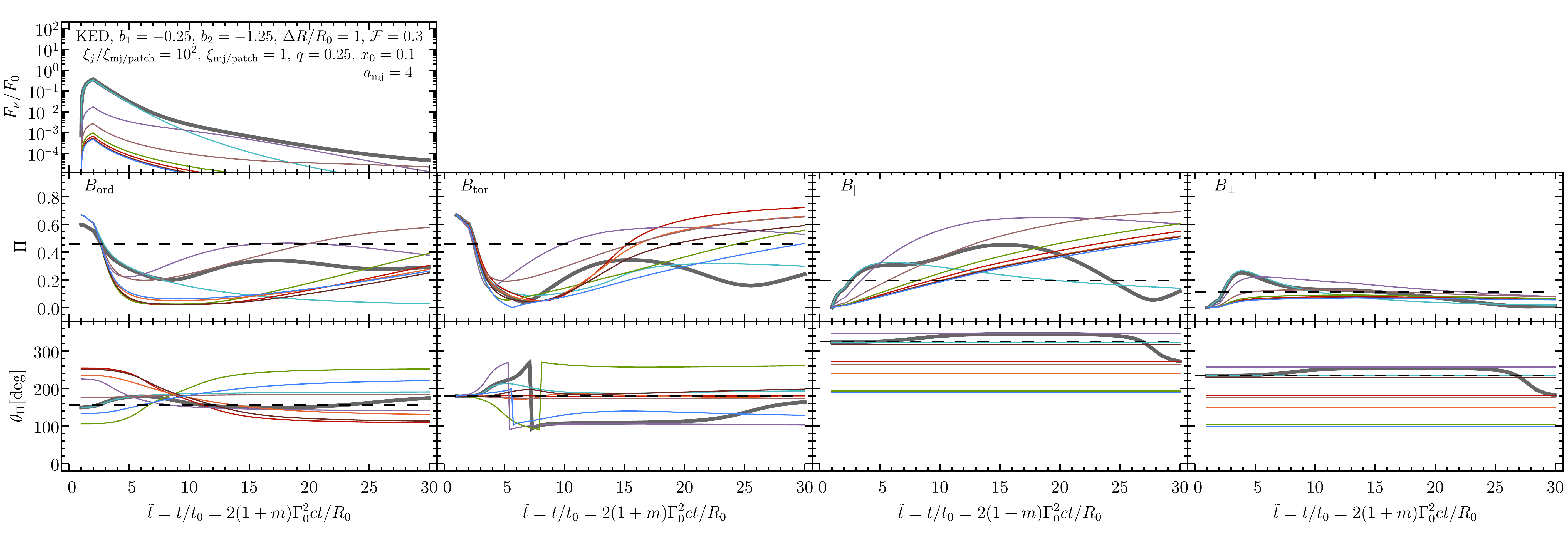}
    \caption{Same as Fig.\,\ref{fig:PLmj-in-TH-jet} but with a steeper ($a_{\rm mj}=4$) emissivity profile of the \MJs.
    }
    \label{fig:PLmj-in-TH-jet-a_mj-4}
\end{figure*}

\begin{figure*}
    \centering
    \includegraphics[width=0.48\textwidth]{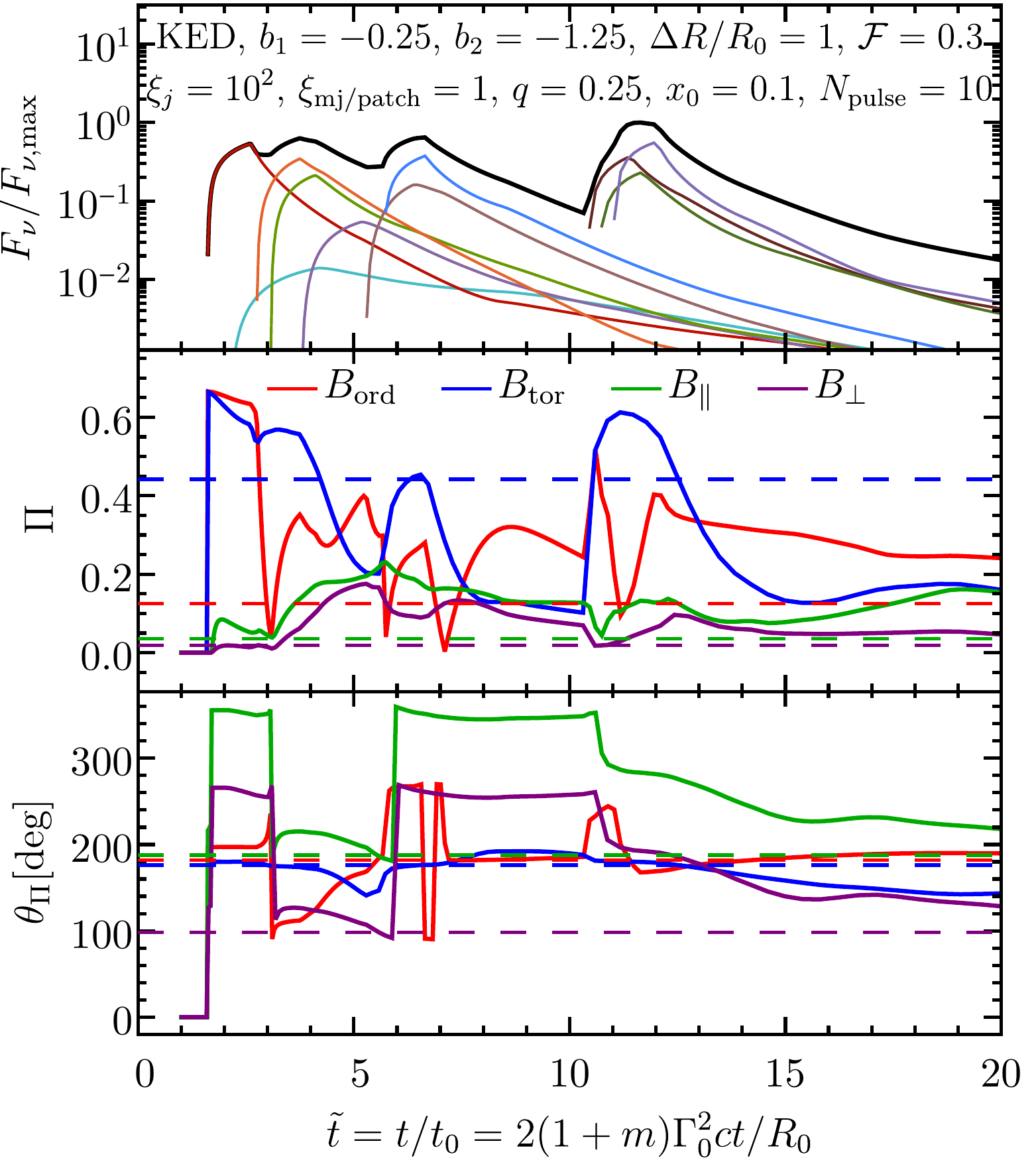}\hspace{2em}
    \includegraphics[width=0.48\textwidth]{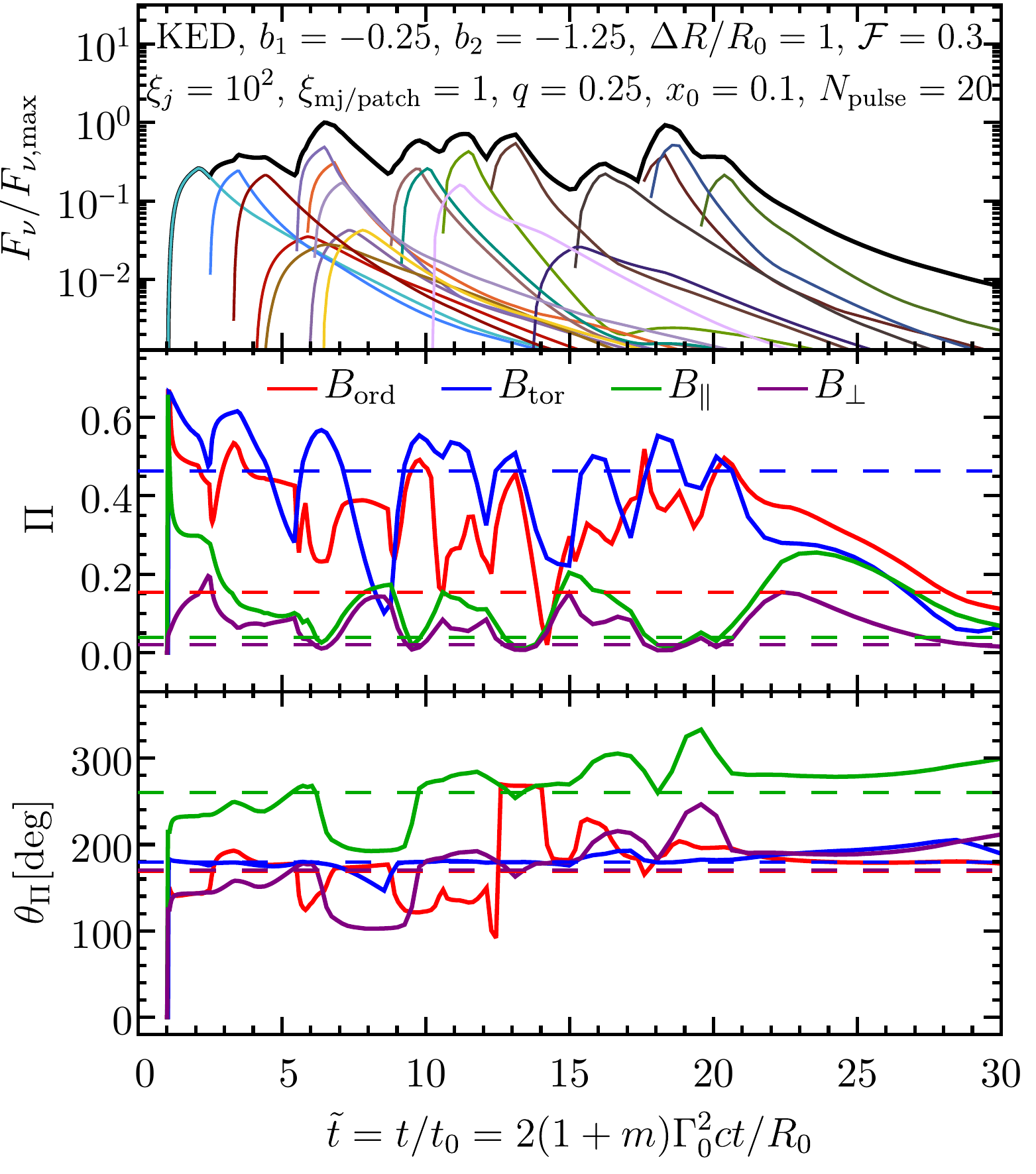}
    \caption{Polarization evolution of multiple (\textbf{Left}: $N_{\rm pulse}=10$, and \textbf{Right}: $N_{\rm pulse}=20$) 
    overlapping pulses, shown for different B-fields. All else is the same as in Fig.\,\ref{fig:THmj-in-THJ-diff-Bfield} 
    (namely top-hat MJs or uniform patches within a top-hat global jet) 
    except for the fact that each pulse is obtained from a different random realization of the distribution of MJs/patches 
    inside the global jet. The dashed lines in the middle panel show the time-integrated polarization.}
    \label{fig:MP-THmj-in-THJ-diff-Bfield}
\end{figure*}

Fig.\,\ref{fig:THmj-in-PL-jet} shows the polarization evolution of several top-hat \MJs~within the aperture of 
a global jet with power-law angular structure in both the emissivity ($a_j=1$) and bulk-$\Gamma$ ($b_j=1$), and for an 
observer with $q\equiv\theta_{\rm obs}/\theta_c=0.25$. 
Polarization trends similar to that shown in Fig.\,\ref{fig:THmj-in-THJ-diff-Bfield} are obtained here. The main difference is that due to the power-law angular 
profile in emissivity, \MJs~within the core make the dominant contribution while those outside of the core remain insignificant. 
In addition, the growing sizes of the \MJs~due to the bulk-$\Gamma$ angular profile limit their total number for a given covering factor. 

Figure\,\ref{fig:THmj-in-PL-jet-q-1.25} shows the effect of a larger viewing angle with $q=1.25$. Here the emission is dominated 
by a single MJ/patch, with $\theta_{\rm patch/mj}>\theta_{\rm obs}$ (but $\bar{\theta}_{\rm patch/mj}<\theta_{\rm obs}$ ), 
closest to the observer's LOS. Even though the MJs/patches at $\theta_{\rm patch/mj}<\theta_{\rm obs}$ that contribute to the emission 
are intrinsically brighter, they still do not make a dominant contribution due to the Doppler de-beaming of their emission caused 
by their larger angular distance away from the LOS and their larger LFs.

Figure\,\ref{fig:THmj-in-PL-jet-a-3-q-2} shows an even more extreme case with $q=2$ and much steeper angular profile in emissivity ($a_j=3$).  
It shows qualitatively the same behavior as was seen in Fig.\,\ref{fig:THmj-in-PL-jet-q-1.25}. This arises since the huge Doppler 
de-beaming of the emission from \MJs~within the jet's core still wins by a large margin over their larger intrinsic brightness even 
for fairly steep jet emissivity angular profiles.

Figure\,\ref{fig:THmj-in-PL-jet-ximj-4} shows the effect of a larger angular size of the MJs/patches in the core, 
or equivalently of larger $\xi_{\rm patches/mj,c}$. This increases the polar angular scale $\theta_{\rm mj\star}$ out to which the 
MJs/patches maintain a fixed angular size before they start to grow. For $\xi_{\rm patches/mj,c}=4$ and $b_j=1$, 
$\theta_{\rm mj\star}/\theta_c=\sqrt{3}$. Due to the fixed angular size of the MJs/patches, this case looks rather similar to 
that shown in Fig.\,\ref{fig:THmj-in-THJ-diff-Bfield-ximj-4}, but the important difference between the two cases is that here 
the bulk $\Gamma$ and emissivity of each MJ/patch changes with the global jet angular profile.

Figure\,\ref{fig:THmj-in-PL-jet-xic-4-q-1-F-0.3} explores an interesting scenario where the global lightcurve now shows two 
separate peaks. This is in contrast to all of the earlier cases that show only a single peaked lightcurve. In a global jet with 
angular structure in both emissivity and bulk $\Gamma$, two effects play an important role in determining the flux of a 
MJ/patch when the LOS of the observer is outside the aperture of that MJ/patch, i.e. $q_{\rm patch/mj}>1$. First, if the 
observer is off-axis w.r.t. the symmetry axis of the global jet, with $q=\theta_{\rm obs}/\theta_c>0$, the MJ/patch at 
$\theta_{\rm patch/mj}<\theta_{\rm obs}$ will be intrinsically brighter than the one at $\theta_{\rm patch/mj}>\theta_{\rm obs}$, 
depending on the exact power-law decay in emissivity with angle $\theta_{\rm patch/mj}$. Second, when $q_{\rm patch/mj}>1$, 
emission from the MJ/patch is Doppler de-beamed, making it dimmer. In the case shown, the MJ/patch that forms the second peak 
in the global lightcurve is much closer to the LOS compared to a few other MJs/patches that have approximately the same 
$\theta_{\rm patch/mj}$ but have a larger angular distance from the LOS.

The effect of different $\xi_{\rm patch/mj,c}$ is shown in Fig.\,\ref{fig:THmj-in-PLJ-diff-Bfield-diff-ximj}. As before, the larger angular 
sizes of the \MJs~curtail the contribution to the total emission from larger number of \MJs, whereby only a single MJ/patch may make 
the dominant contribution in most cases. This leads to less chaotic changes in the PA where it remains steady over longer 
fractions of the pulse duration. 

When the global jet has a much steeper power-law profile of bulk-$\Gamma$, the issue of having only a few \MJs~contribute to the 
total emission becomes even more pronounced. In the example shown in Fig.\,\ref{fig:THmj-in-PLJ-diff-Bfield-diff-bPLJ}, as $b_j$ 
becomes larger (steeper profile) only those MJs/patches closer to the LOS start to dominate over the entire pulse.

The effect of different jet core sizes is explored in Fig.\,\ref{fig:THmj-in-PLJ-diff-Bfield-diff-xic}, where smaller core sizes, 
corresponding to smaller $\xi_c$ values, means that for a given covering factor only a small number of MJs can be introduced 
within $\theta_{\max}=2\theta_c$. It also means that the angular sizes of the MJs will start to grow much closer to the jet 
symmetry axis.

\subsection{Angular Structured MJs/Patches Inside a Top-Hat Jet}
The angular structure of the MJs is not limited to the highly idealized top-hat profile. Therefore, we consider here MJs with 
a power-law angular structure inside a uniform global jet. The angular structure is considered only for the comoving emissivity 
alone, with a uniform angular profile in $\Gamma_{\rm mj}$, such that 
\begin{equation}
\nu_p'L'_{\nu_p',\rm mj}=L'_{0,\rm mj}\Theta_{\rm mj}^{-a_{\rm mj}}\ ,\quad\quad \Theta_{\rm mj}(\bar\theta)=\sqrt{1+(\bar\theta/\bar\theta_{\rm mj})^2}\,,
\end{equation}
where the angular profile depends on $\bar\theta$, which is the polar angle measured from the symmetry axis of each MJ, 
whose core angle is $\bar\theta_{\rm mj}$. The luminosity normalization $L_{0,\rm mj}'= \nu_p'L_{\nu_p',{\rm mj}}'(\bar\theta = 0)$ 
is the same for all of the MJs. 

Here we do allow for the possibility for the patches to also have a similar angular structure in emissivity. This is mainly 
done for comparison between the different B-field cases. In reality, large emissivity gradients should be short-lived after a shell 
collision and the patches must become approximately uniform in a region of angular size $\sim 1/\Gamma$ over a single dynamical time. 
On the other hand, the emission may last only over a single dynamical time, during which significant gradients are not yet efficiently 
washed out.

The top-panel of Fig.\,\ref{fig:PLmj-in-TH-jet} shows the distribution of MJs/patches, where the non-overlapping condition is set to having 
the centers of the two \MJs~be apart by at least $4\bar\theta_{\rm mj}$. 
Since there is always material emitting along the LOS when the \MJs~have angular structure, the pulse 
onset is without any offset and starts at $\tilde t=1$, even from the distant ones. In the shown setup, the MJ/patch centered 
closest to the LOS makes the dominant contribution over the distant ones. In the $B_{\rm ord}$ case, the global polarization 
is reduced due to partial cancellation owing to contributions to the emission by other \MJs~with different PA, even though 
their contribution is subdominant. 
A similar reduction in the initial global $\Pi$ does not happen for $B_{\rm tor}$, at least until $\tilde t\sim10$, since the 
PAs of all the dominantly contributing patches are rather similar, which avoids cancellation of the polarization.
In the $B_\parallel$ and $B_\perp$ cases, the initial polarization at $1 \leq \tilde t\leq 2$ is very small due to 
almost complete cancellation of the polarization vectors owing to the fact that within the beaming cone the prescribed angular 
structure is insufficient to strongly break the symmetry around the LOS. However, after the pulse-peak at $\tilde t > 2$, when 
high-latitude emission starts to become important, the polarization starts to grow and reaches a mean value before declining again. 
In all cases, the evolution of the PA tracks that of the MJ/patch that makes the dominant contribution to the total emission. 

Figure\,\ref{fig:PLmj-in-TH-jet-a_mj-4} shows the effect of a much steeper ($a_{\rm mj}=4$) emissivity angular 
profile of the \MJs. The greater variation of the emissivity within the beaming cone offers a way to increase the polarization 
of the \MJs~that are closest to the LOS and also contribute dominantly. As a result, the global polarization also 
increases. A second effect of having a steeper profile is the enhancement in the contribution of the MJ/patch closest to the 
LOS over the others. This results in reduced cancellations in the Stokes plane and thus a higher global polarization. Finally, 
neighboring \MJs~can dominate the total flux during the tail of the pulse leading to secondary peaks in the polarization and 
larger changes in the PA.

Figure\,\ref{fig:PLmj-in-THJ-diff-Bfield-diff-xicmj} shows the variation of $\Pi$ and $\theta_{\Pi}$  
over the pulse profile for different values of $\xi_{\rm mj}=(\Gamma\bar\theta_{\rm mj})^2$. 
To keep the same number of MJs while $\xi_{\rm mj}$ is increased we fix the ratio $\xi_j/\xi_{\rm mj}=(\theta_j/\bar\theta_{\rm mj})^2=10^2$. 
Larger levels of polarization over the pulse is obtained when $\xi_{\rm mj}$ is smaller. This can be simply understood by noticing 
that for a smaller $\xi_{\rm mj}$, the angular size of the observed region 
$1/\Gamma=\xi_{c,\rm mj}^{-1/2}\theta_{\rm mj}$ is a larger fraction of the core angle, which then permits a larger variation 
of the emissivity within the observed region for a given $a_{\rm mj}$. As $\xi_{\rm mj}$ is increased the angular size of the 
observed region shrinks w.r.t the core angle, which leads to smaller variations of the emissivity in the observed region, 
and therefore smaller levels of $\Pi$. Correspondingly, the change in the PA also diminishes.

\section{Multiple Overlapping Pulses}\label{sec:multi-pulse}
In the earlier sections, we only considered a single pulse that would result from an isolated energy dissipation 
episode, e.g. a single collision between two shells in the internal shocks scenario. However, such cases are 
rare and typically prompt GRB emission shows multiple overlapping pulses with a variety of pulse widths. The onset 
time of the $i^{\rm th}$ pulse, e.g., in the internal shocks scenario is given by 
\begin{equation}
    t_{{\rm onset},z,i} = t_{{\rm ej},z,i} + t_{0,z,i} = t_{{\rm ej},z,i} + \frac{R_{0,i}}{2(1+m)\Gamma_{0,i}^2c}\,,
\end{equation}
where $t_{{\rm ej},z,i}$ is the ejection time of the shell and $t_{0,z,i}$ is the radial delay time. This equation is strictly true for spherical shells or if there is emitting material along the LOS. Otherwise, the 
arrival time of the initial photons incurs an additional angular time delay, as shown in Eq.\,(\ref{eq:EATS}).
In addition to the above, emission from each shell collision can last over 
different $(\Delta R)_i/R_{0,i}$ radial distances, giving rise to different pulse durations. In the Poynting-flux-dominated 
case, the ejection time can be replaced by the timescale over which an MHD disturbance occurs. To keep the treatment 
simpler, here we dispense with the details of the shell energization process that produces multiple pulses, and 
instead shift the onset time of each pulse by $\Delta t_{{\rm onset},z,i}$ with respect to the onset time of the first 
pulse. This time shift is drawn randomly from a uniform distribution with $0\leq\Delta t_{{\rm onset},z,i}\leq\Delta t_{\max}$.

In Figure\,\ref{fig:MP-THmj-in-THJ-diff-Bfield} we show the polarization evolution with B-field configurations 
for $N_{\rm pulse}=\{10,20\}$ overlapping pulses with $\Delta t_{\max}/t_0=\{10, 20\}$ and $(\Delta R)_i/R_{0,i}=1$. 
The rest of the setup is the same as in Fig.\,\ref{fig:THmj-in-THJ-diff-Bfield}, but each pulse constitutes a different 
random realization of the distribution of the \MJs~inside the global jet. The polarization curve shows multiple peaks 
that display a weak correlation with the peaks in the lightcurve. Since there could be multiple overlapping dimmer pulses, 
addition of their Stokes parameters might enhance or diminish the net polarization and change the PA accordingly. 

The time-integrated polarization, shown as horizontal dashed lines, is the highest for the $B_{\rm tor}$ field, due to its modest 
variation of the time-resolved PA. When integrated over the duration of the emission episode, such a steady PA results in 
very little cancellations in the Stokes $Q-U$ plane, thus yielding high time-integrated polarization with $\Pi\sim45$ per cent. 
In contrast, even though the $B_{\rm ord}$ field also features a large-scale ordered field within each MJ/patch, the extra degree of freedom of having a 
randomly oriented B-field results in greater variation of the PA over the emission episode. This leads to a significantly reduced 
time-integrated polarization, with $\Pi\sim15-20$ per cent, due to cancellations in the Stokes $Q-U$ plane. 
Likewise for the $B_\perp$ and $B_\parallel$ fields, 
since the plane of net PA from each MJ is randomly oriented, the large variation in the global PA among different pulses again tends 
to cancel out the polarization and yield only small time-integrated polarization at the level of a few per cent. Furthermore, the 
time-integrated polarization remains approximately the same for all the B-field configurations as the number of pulses are doubled.
A caveat is that we have only shown two random realizations of the whole process of 
generating $N_{\rm pulse}$ pulses, and a different random realization may show a somewhat different temporal evolution of polarization. 
However, the time-integrated polarization is expected to be more robust and remain steady for large $N_{\rm pulse}$.

\section{Model parameters: what can be constrained?}
The model described in this work entails a fairly large number of parameters (see Table\,\ref{tab:symbols}) that describe the dynamical, 
spectro-polarimetric, and structural features of the global jet as well as the MJs/patches. While being comprehensive in addressing 
the different effects caused by the different features, this parameterized model also affords a fair degree of flexibility when 
compared with observations. Therefore, it is not advisable to try to constrain all of the model parameters using a given observation, 
as the parameter space is degenerate. Instead, we recommend to only constrain a few of the most important model parameters, as allowed by the effective number of constraints from the data, while keeping the others fixed. Based on the type of observation, different model parameters can be constrained:
\begin{enumerate}
    \item \textit{Pulse profile}: The pulse profile of emission from a single MJ/patch at a fixed normalized energy $x_0 = \nu/\nu_0$, 
    or integrated over a given energy bin $\Delta\nu/\nu_0$, is most sensitive to the jet dynamics, e.g. the PL indices $a$ and $d$ 
    that describe the radial profile of the comoving spectral emissivity, $\Delta R/R_0$, and $m$ that gives the acceleration profile 
    of the global jet. When multiple MJs/patches contribute to the emission, that now also features a steep to shallow trend, the location and covering 
    factor, where the latter sets the density of MJs/patches near the LOS, becomes important. Evidence for multiple MJs/patches can 
    actually be obtained in bright single pulse GRBs or those that show isolated broad pulses. There is some degeneracy between the 
    covering factor, location, and angular size ($\xi_{\rm patch/mj}$) of the MJs/patches when describing the pulse profile.
    \item \textit{Spectrum}: The spectrum over a given energy range is used to constrain the two spectral indices $b_1$ and $b_2$. 
    These are important for determining the absolute maximum local polarization for synchrotron emission.
    \item \textit{Polarization}: Both the time-resolved and time-integrated polarization are sensitive to the magnetic field configuration, 
    albeit some degeneracy between the different cases considered here is still expected. In most cases the ordered fields typically 
    yield $\Pi\gtrsim20$\,per cent. This rough dividing line can be used to separate out the ordered fields from the small-scale shock-produced 
    axisymmetric fields.
\end{enumerate}

The best constraints are achieved in a joint time-resolved pulse-profile and spectro-polarimetric fit that offers the most number of constraints. 
One sensible way to learn about the jet properties is to either assume a KED or PFD flow, which will fix $a$, $d$, and $m$ a priori. Then, 
one can choose to interpret the observations using either a uniform jet or that with angular structure.

All of the model degeneracies that affect the single pulse case are naturally more pronounced when multiple overlapping pulses contribute to the 
emission. However, the time-integrated polarization and PA in this case can yield the \textit{smoking-gun} evidence for a large-scale ordered 
B-field, e.g. a globally ordered toroidal field. Furthermore, the distinction between the $B_{\rm ord}$ and $B_{\rm tor}$ fields can only be 
made more robustly with mulitiple overlapping pulses and not so much in a single pulse GRB.

Finally, all of the results shown in this work are based on a single random realization of the distribution of MJs/patches. A different 
distribution will yield slightly different results (as shown in Fig.\,\ref{fig:THmj-in-THJ-diff-Bfield-diff-seed}). 
However, it is expected that different random realizations will statistically yield 
broadly similar trends of pulse profiles and polarization evolution. Therefore, to be more prudent, one should try a number of random 
realizations for any given scenario to test the robustness of the fit.

\section{Conclusions \& Discussion}\label{sec:conclusion-discussion}

In this work, we have extended the treatment in \citet{Gill-Granot-21} of time-resolved polarization from different 
B-field configurations. The introduction of multiple radially expanding non-axisymmetric MJs or emissivity 
patches lead to a gradual and continuous change of the PA. Some notable features that have emerged in the different 
scenarios explored here are:
\vspace{-0.1cm}
\begin{enumerate}[leftmargin=0cm,itemindent=.5cm,itemsep=0.6em,labelwidth=\itemindent,labelsep=0cm,align=left,label=({\bf\roman*})]
\item \textit{Continuous change in the PA}: A continuous change in the PA is the most 
important outcome of the MJ/patches model and it is obtained in all B-field configurations considered here. 
Such behavior cannot be obtained in axisymmetric jets. 

\item \textit{Polarization dilution}: When several MJs/patches contribute to the flux, the incoherent nature 
of the emission leads to a random walk in the Stokes $Q_\nu$ and $U_\nu$ parameters space. As a result, the net 
polarization is diluted to $\Pi\sim\Pi_0/\sqrt{N}$ where $N$ are the number of MJs/patches observed (i.e. effectively 
contribuing to the emission) at any given time 
and $\Pi_0$ is the polarization from a single MJ/patch that makes the dominant contribution to the flux. This becomes 
important for time-resolved polarization model fitting in which emission from a uniform flow will yield a 
significantly different polarization evolution as compared to one with multiple MJs/patches. In particular, the polarization from a large-scale ordered field, e.g. $B_{\rm ord}$ and $B_{\rm tor}$, in a 
uniform flow (or when $q=\theta_{\rm obs}/\theta_j\ll1$), can also decline in the tail of the pulse, but the PA remains constant. This feature serves as a good diagnostic that can distinguish between emission from a uniform and non-axisymmetric 
flow containing multiple MJs/patches. In addition, the decay of both the lightcurve and polarization in a uniform flow is expected to be much smoother in comparison to that obtained from multiple MJs/patches.

\item \textit{Time-integrated polarization}: The single-pulse time-integrated polarization remains consistently higher 
in a scenario consisting an ordered B-field ($B_{\rm ord}$ and $B_{\rm tor}$) in comparison to a small-scale shock-produced field 
($B_\perp$ and $B_\parallel$). This feature is similar to what is also found in uniform jet models with $q<1$.

\item \textit{Steep-to-shallow pulse profiles}: In uniform jet models with $q<1$, the pulse profile of a single pulse 
shows a power-law decline after the emission from the shell terminates and is dominated by the high-latitude emission. 
With the addition of two jet breaks, the pulse profile can only become steeper \citep[see, e.g., Fig.\,7 of][]{Gill-Granot-21}. 
In the MJ/patchy-shell scenario, a steeply declining lightcurve can become shallow due to contributions from other 
\MJs~outside of the beaming cone whose emission peaks at a later time. Therefore, this represents one possible way of 
obtaining a steep to shallow behavior in a given energy band across a temporal break in some GRBs that show a single-pulse 
prompt emission. 
    
\item \textit{Distinguishing features of an ordered B-field}: Since an ordered B-field necessarily breaks the symmetry, at least locally, the polarization is high at the start of the pulse, regardless of the viewing geometry. 
Comparably high initial polarization is also obtained for $B_\parallel$ and $B_\perp$ but only when 
$q_{\rm mj}\equiv\bar\theta/\bar\theta_{\rm mj}>1$, i.e. the MJ/patch is viewed from outside of its own aperture.
    
\item \,\textit{Structured \MJs}: When the \MJs~have angular structure in emissivity, the difference between large-scale ordered fields 
($B_{\rm ord}$ and $B_{\rm tor}$) and the small-scale shock-produced fields ($B_\perp$ and $B_\parallel$) becomes even larger and more 
readily apparent. An ordered field will always yield very high polarization during the rising phase of the pulse, whereas the 
$B_\parallel$ and $B_\perp$ cases will always yield nearly negligible polarization during this time. Furthermore, the angular structure 
plays an important role, and a steeper profile yields a higher global (time- resolved and integrated) polarization over a shallower one, 
regardless of the B-field configuration.
    
\item\ \textit{Multiple overlapping pulses}: The level of time-resolved polarization for the $B_\parallel$ and $B_\perp$ cases 
is significantly reduced for multiple overlapping pulses as compared to a single pulse. In contrast, the level of polarization 
remains high for a large-scale ordered field. With the exception of $B_{\rm tor}$, the time-integrated polarization of an emission 
episode with multiple overlapping pulses is significantly reduced for $B_{\rm ord}$, $B_\parallel$, and $B_\perp$ fields due to 
larger time variations of their global PAs. Since the $B_{\rm tor}$ field is axisymmetric about the jet symmetry axis, the 
PA remains approximately steady, which produces a high time-integrated polarization. Therefore, $B_{\rm tor}$ is the only 
field geometry out of the four considered here that can consistently yield time-integrated $\Pi\gtrsim40$ per cent. The same conclusion was 
reached by \citet{Gill+20} who carried out a statistical study using different B-field configurations and axisymmetric jet structures.
\end{enumerate}

The time-resolved analysis of a single-pulse GRB\,170114A revealed a trend of growing polarization over the rising 
phase of the pulse where it reached $\Pi\sim30\%$ \citep{Zhang+19,Burgess+19}. During this time a continuous and gradual change 
in the PA was also noted, due to which the time-integrated analysis over the entire duration of the pulse found a low 
level of polarization with $\Pi\sim4\%$. When interpreting these findings using the MJ/patches model, any ordered field 
scenario is ruled out since in this case initially large polarization that declines during the rising phase of the pulse 
is expected. Furthermore, the single pulse time-integrated polarization for an ordered field tends to be much higher 
($\Pi\gtrsim40$ per cent) than what was measured. These observations are most consistent with the shock-produced small-scale 
field ($B_\perp$ and $B_\parallel$) scenarios that do not show an initial spike to large polarization when $q_{\rm mj}<1$, 
and may only reach a time-resolved polarization of at most $\Pi\sim40$ per cent in many cases. Their single-pulse time-integrated 
polarization can also be modest with $\Pi\sim$ few per cent in some cases. Some of these features are in fact similar to what was seen for GRB\,170114A. 
We caution the reader, though, that the modest statistical significance of these observation does not allow to draw strong conclusions from them. 

The time-binned analysis of GRB\,100826A, observed by IKAROS-GAP, claimed a firm change in the PA at the $\sim3.5\sigma$ level 
between two $50\,$s time intervals that comprise mulitple overlapping pulses \citep{Yonetoku+11}. Best 
fit values of $\Pi_1=25\%\pm15\%$ and $\Pi_2=31\%\pm21\%$ with PA of $\theta_{\Pi,1}=159\pm18$ and $\theta_{\Pi,2}=75\pm20$ 
Given the lower significance of the detection with large uncertainties, it is difficult to rule out any B-field 
configuration discussed here. Except for the fact that, at face value, the measurements do find a significant difference between 
the time-integrated PAs of the two emission episode. This feature cannot be accommodated by the $B_{\rm tor}$ field for which the 
time-resolved and time-integrated PA of multiple overlapping pulses is along the line connecting the jet symmetry axis and the observer's LOS.

A continuously evolving PA is also obtained in time-resolved non-axisymmetric photospheric models of prompt emission. This is demonstrated in 
the recent work by \citet{Ito+23} where they use three-dimensional hydrodynamic simulations for outflow properties and then 
post-process the numerical results with Monte-Carlo radiation transfer simulations to calculate the time-resolved Stokes 
parameters. Earlier works \citep[e.g.][]{Parsotan+20,Ito+21,Parsotan-Lazzati-22} used a similar numerical technique to calculate 
time-resolved polarization but used two-dimensional simulations. This made the outflow geometry axisymmetric and, therefore, 
restricted changes in the PA to only $90^\circ$.

To gain a better understanding of the prompt GRB radiation mechanism and non-axisymmetric structure of the flow, we have to 
wait for the launch of more sensitive and dedicated GRB polarimeters, namely POLAR-2 \citep{deAngelis+22} and LEAP \citep{McConnell+16}, 
that can finally provide highly statistically significant measurements. Once they become available, models like the one presented 
in this work will have enough complexity to fit to observations and deliver robust results. 

\section*{Acknowledgements}
R.G. acknowledges financial support from the UNAM-DGAPA-PAPIIT IA105823 grant, Mexico. 
J.G. acknowledges financial support by the ISF-NSFC joint research program (grant no. 3296/19).

\section*{Data Availability}
Data from the numerical model can be shared on reasonable request.


\bibliographystyle{mnras}
\bibliography{refs} 

\appendix
\section{Additional Figures}
\label{sec:appendix}

\begin{figure*}
    \centering
    \includegraphics[width=0.98\textwidth]{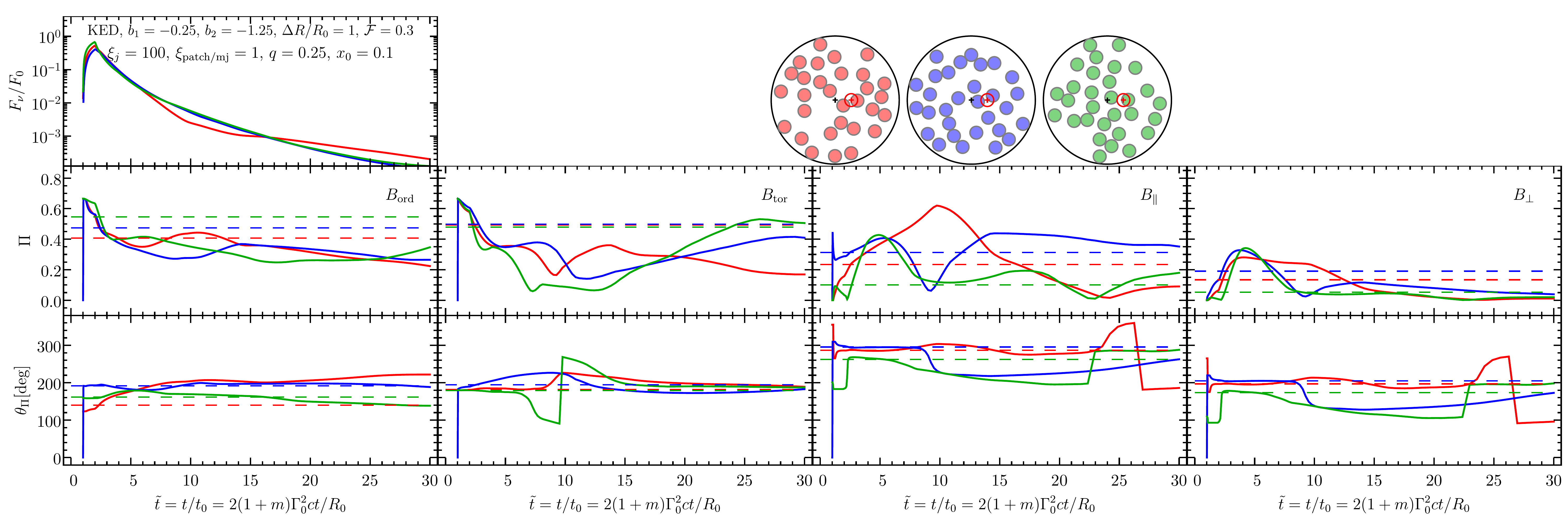}
    \caption{Same as Fig.\,\ref{fig:THmj-in-THJ-diff-Bfield} but with three different random realizations of the distribution of \MJs.
    }
    \label{fig:THmj-in-THJ-diff-Bfield-diff-seed}
\end{figure*}

\begin{figure*}
    \centering
    \includegraphics[width=0.98\textwidth]{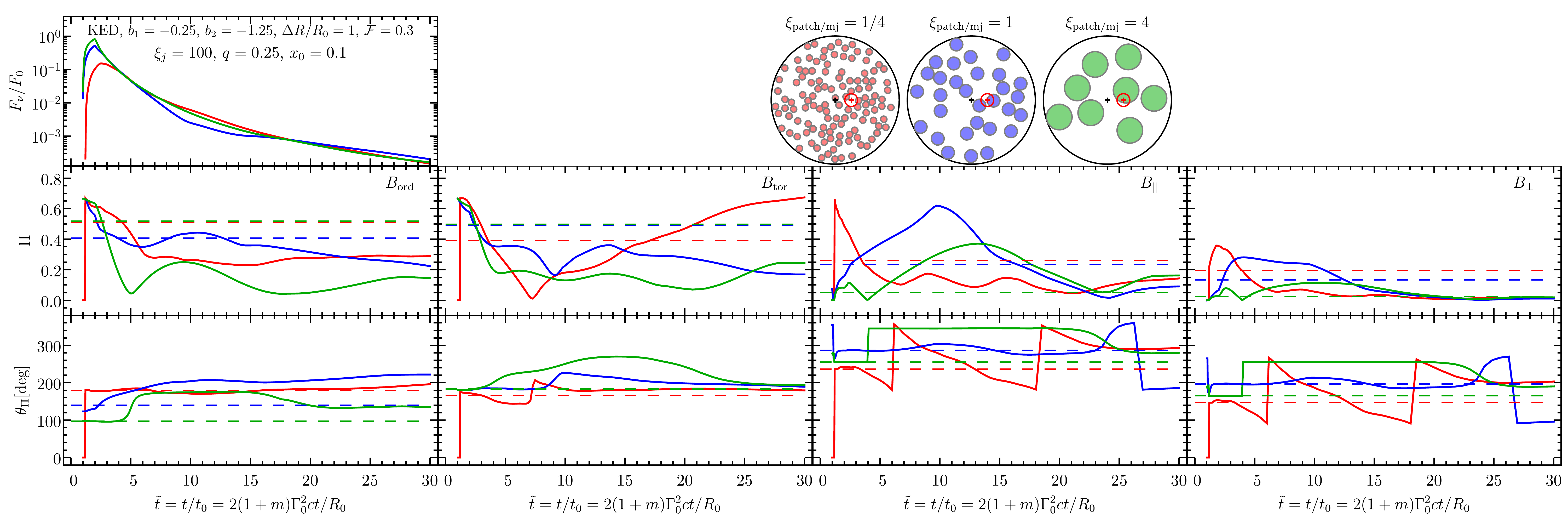}
    \caption{Same as in Fig.\,\ref{fig:THmj-in-THJ-diff-Bfield} but with different $\xi_{\rm patch/mj}$ parameters.
    }
    \label{fig:THmj-in-THJ-diff-Bfield-diff-ximj}
\end{figure*}

\begin{figure*}
    \centering
    \includegraphics[width=0.98\textwidth]{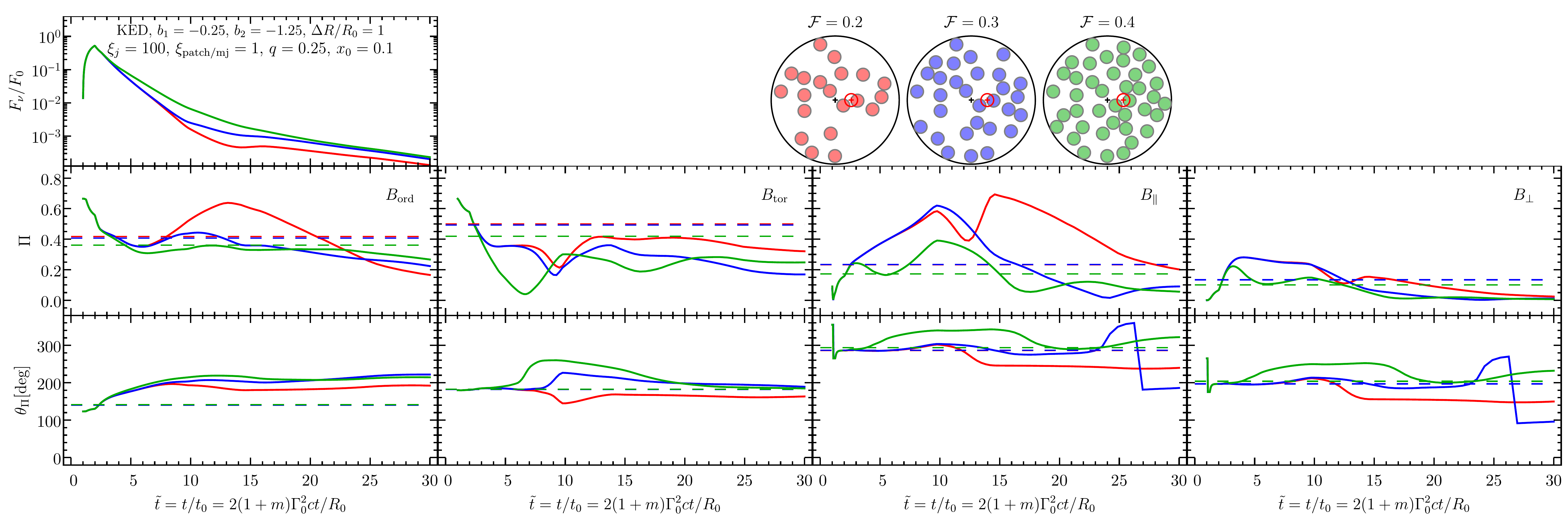}
    \caption{Same as Fig.\,\ref{fig:THmj-in-THJ-diff-Bfield} but with different covering factors $\mathcal{F}$. 
    As $\mathcal{F}$ (and correspondingly $N_{\rm patch/mj}$) is increased, new MJs/patches drawn from the same 
    distribution are added to the previous ones.
    }
    \label{fig:THmj-in-THJ-diff-Bfield-diff-Fcover}
\end{figure*}

\begin{figure*}
    \centering
    \includegraphics[width=0.98\textwidth]{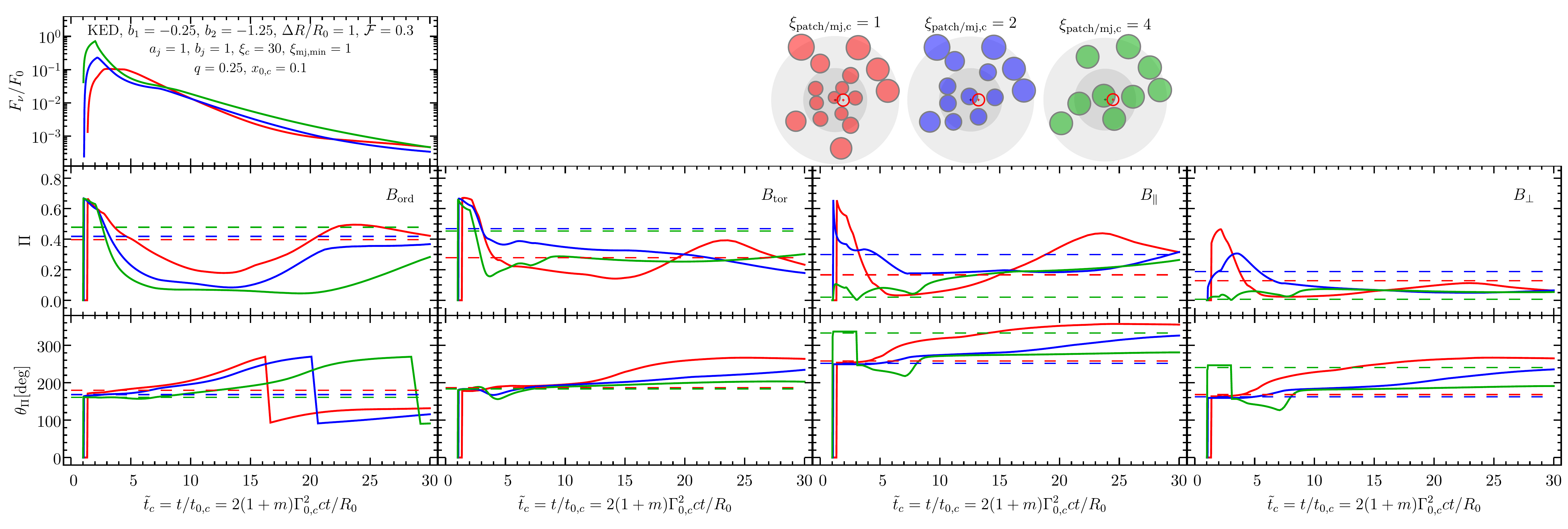}
    \caption{Same as Fig.\,\ref{fig:THmj-in-PL-jet} but with different $\xi_{\rm patch/mj,c}$.
    }
    \label{fig:THmj-in-PLJ-diff-Bfield-diff-ximj}
\end{figure*}

\begin{figure*}
    \centering
    \includegraphics[width=0.98\textwidth]{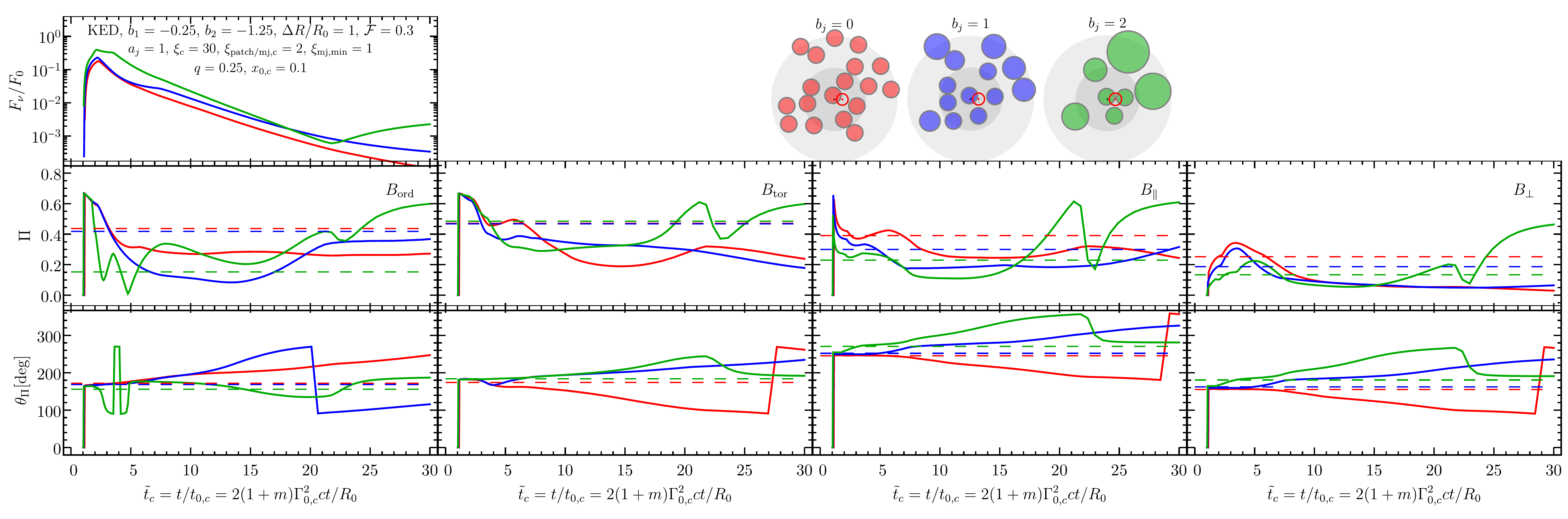}
    \caption{Same as Fig.\,\ref{fig:THmj-in-PL-jet} but for $\xi_{\rm patch/mj,c}=2$ and with different $b_j$ values.
    }
    \label{fig:THmj-in-PLJ-diff-Bfield-diff-bPLJ}
\end{figure*}

\begin{figure*}
    \centering
    \includegraphics[width=0.98\textwidth]{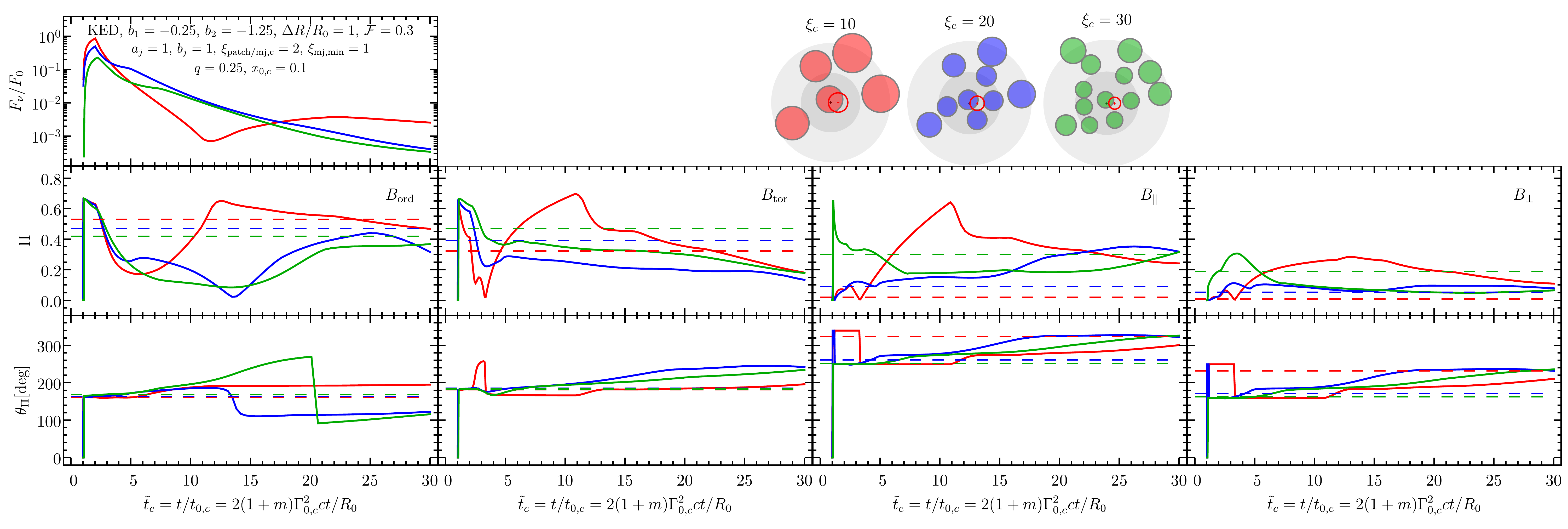}
    \caption{Same as Fig.\,\ref{fig:THmj-in-PL-jet} but with $\xi_{\rm patch/mj,c}=2$ and different $\xi_c$ values.
    }
    \label{fig:THmj-in-PLJ-diff-Bfield-diff-xic}
\end{figure*}

\begin{figure*}
    \centering
    \includegraphics[width=0.98\textwidth]{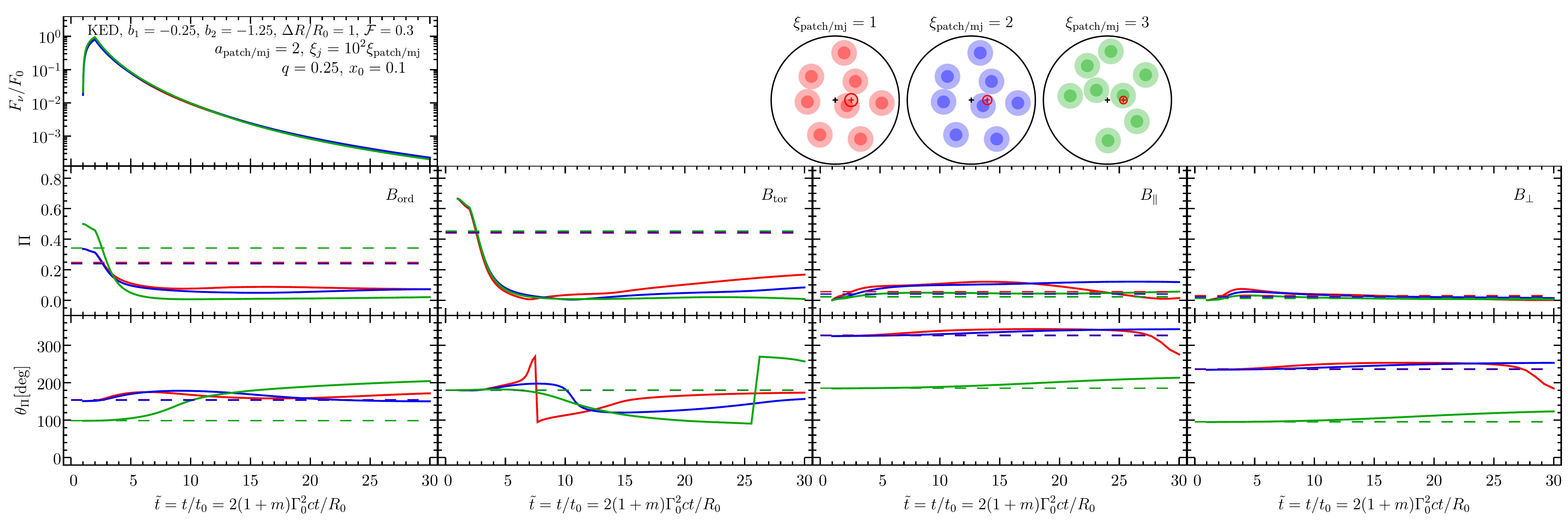}
    \caption{Same as Fig.\,\ref{fig:PLmj-in-TH-jet} but with different values of $\xi_{\rm patch/mj}=\xi_j/100$.
    }
    \label{fig:PLmj-in-THJ-diff-Bfield-diff-xicmj}
\end{figure*}



\bsp	
\label{lastpage}
\end{document}